\documentclass{article}

\usepackage[preprint]{neurips_2025}

\usepackage[utf8]{inputenc} 
\usepackage[T1]{fontenc}    
\usepackage{hyperref}       
\usepackage{url}            
\usepackage{booktabs}       
\usepackage{amsfonts}       
\usepackage{nicefrac}       
\usepackage{microtype}      
\usepackage{xcolor}         
\definecolor{mydarkgreen}{RGB}{0,200,0}
\definecolor{mydarkred}{RGB}{200,0,0}
\usepackage{graphicx}

\usepackage{subfigure}
\usepackage{array} 
\usepackage{adjustbox}
\usepackage{amsmath}
\usepackage{amssymb}
\usepackage{physics} 
\usepackage{siunitx} 
\usepackage{multirow}
\usepackage{soul}

\usepackage{bm}
\usepackage{caption}
\usepackage{subcaption}

\usepackage{titletoc}

\usepackage{etoolbox}
\newcommand{\bcheck}{\textcolor{mydarkgreen}{\textbf{$\checkmark$}}}
\newcommand{\btimes}{\textcolor{mydarkred}{\textbf{$\times$}}}

\title{TripNet: Learning Large-scale High-fidelity 3D Car Aerodynamics with Triplane Networks}

\author{%
  Qian Chen\thanks{Corresponding author} \\
  Department of Mechanical Engineering\\
  Massachusetts Institute of Technology\\
  Cambridge, MA 02139 USA \\
  \texttt{chenqian@mit.edu} \\
  \And
  Mohamed Elrefaie \\
  Department of Mechanical Engineering\\
  Massachusetts Institute of Technology\\
  Cambridge, MA 02139 USA \\
  mohamed.elrefaie@mit.edu \\
  \And
  Angela Dai \\
  Department of Computer Science \\
  Technical University of Munich\\
  Garching, 85748 Germany\\
  angela.dai@tum.de \\
  \And
  Faez Ahmed \\
  Department of Mechanical Engineering\\
  Massachusetts Institute of Technology\\
  Cambridge, MA 02139 USA \\
  faez@mit.edu \\
}

\begin{document}

\maketitle

\begin{abstract}
Surrogate modeling has emerged as a powerful tool to accelerate Computational Fluid Dynamics (CFD) simulations. Existing 3D geometric learning models based on point clouds, voxels, meshes, or graphs depend on explicit geometric representations that are memory-intensive and resolution-limited. For large-scale simulations with millions of nodes and cells, existing models require aggressive downsampling due to their dependence on mesh resolution, resulting in degraded accuracy. We present TripNet, a triplane-based neural framework that implicitly encodes 3D geometry into a compact, continuous feature map with fixed dimension. Unlike mesh-dependent approaches, TripNet scales to high-resolution simulations without increasing memory cost, and enables CFD predictions at arbitrary spatial locations in a query-based fashion, independent of mesh connectivity or predefined nodes. TripNet achieves state-of-the-art performance on the DrivAerNet and DrivAerNet++ datasets, accurately predicting drag coefficients, surface pressure, and full 3D flow fields. With a unified triplane backbone supporting multiple simulation tasks, TripNet offers a scalable, accurate, and efficient alternative to traditional CFD solvers and existing surrogate models.
\end{abstract}

\section{Introduction}
\label{sec:intorduction}

Computational simulations have become a cornerstone of engineering and scientific research in recent decades, allowing for the investigation and study of complex physics efficiently and cost-effectively. These simulations solve the partial differential equations (PDEs) to model real-world behavior. Mesh-based simulation approaches such as the finite volume method (FVM) ~\cite{blazek2015computational,patankar2018numerical}, finite element method (FEM) ~\cite{bathe2006finite, belytschko2014nonlinear}, and finite difference method (FDM) ~\cite{pletcher2012computational} have been widely adopted in fields such as aerospace and automotive engineering for computational fluid dynamics and crash simulations. These methods significantly reduce the need for expensive and time-consuming wind tunnel experiments and collision tests.

Simulations are pivotal in driving advancements in science and engineering. Yet, executing large-scale, high-fidelity simulations demands considerable computational resources and time, often extending from several days to weeks per simulation.
To address this issue and accelerate mesh-based simulations, machine learning has emerged as a transformative tool to accelerate traditional simulations by offering faster approximations without sacrificing accuracy. 
Among the promising machine learning approaches, MeshGraphNets~\cite{pfaff2020learning,nabian2024x} have shown considerable potential. These models leverage graph networks to learn high-dimensional scientific simulation data.  The model undergoes training with simulation data through several rounds of message-passing along mesh-edges and world-edges, enabling it to accurately predict dynamics across aerodynamics, structural mechanics, and cloth simulations. However, the scalability of MeshGraphNets is somewhat limited to a few thousand nodes due to the high computational costs associated with building graphs for high-resolution meshes. While MeshGraphNets and similar models~\cite{li2024geometry} have demonstrated significant progress, they face challenges in scaling to large datasets and high-resolution meshes due to computational and memory constraints. Furthermore, alternative representations such as images~\cite{song2023surrogate, elrefaie2024site}, point clouds~\cite{qi2017pointnet, qi2017pointnet++}, and parametric models~\cite{elrefaie2024surrogate} often suffer from either a loss of fine geometric details or excessive computational overhead. Related work is presented in the appendix due to page limitation.

To overcome these challenges, we propose a 3D geometric deep learning model that leverages the triplane representation for tasks such as aerodynamic drag, surface pressure, and 3D flow field prediction.
 Triplanes offer a compact yet expressive structure that is particularly well-suited for CFD tasks. This is because aerodynamic quantities in CFD—such as pressure, wall shear stress, and velocity fields—often exhibit strong anisotropy and boundary-layer behavior aligned with surfaces. Triplane representations naturally encode such surface-oriented features across three orthogonal planes, enabling efficient learning of spatially distributed physical fields without requiring dense 3D volumes or explicit mesh connectivity. This structure strikes a balance between resolution, memory efficiency, and spatial context, making it especially advantageous for high-fidelity CFD prediction. Moreover, since triplanes are extensively employed by generative model~\cite{shue20233d,sun2023next3d, zou2024triplane}, the surrogate models we developed can be seamlessly integrated into various generative AI frameworks for performance-aware 3D generation.

Our contributions are the following:

\begin{enumerate}
    \item We introduce TripNets, a triplane representation method for large-scale, high-fidelity CFD simulation applications. Unlike graph- and point cloud-based models, which are inherently discrete and provide solutions only at the mesh nodes, our method allows the solution to be queried at any point in the 3D space.

    \item TripNets enable aerodynamic predictions for industry-standard cars with remarkable speed and accuracy. Using a single Nvidia RTX 4090 GPU, the method accelerates CFD simulations by several orders of magnitude, reducing the simulation time from 10 hours to seconds. Key performance highlights include drag coefficient tasks less than 0.01 seconds, surface field prediction for 0.5 million nodes under 0.1 seconds, and full 3D velocity field prediction (12 million cells) in 2 seconds compared to traditional CFD simulation.\footnote{For a steady-state RANS simulation on an HPC cluster with 256 CPU cores for 7,000 iterations.} TripNet outperforms prior SOTA models in accuracy on a variety of simulation tasks while being more memory efficient and faster than prior SOTA models: TripNet outperforms all prior models in drag coefficient prediction, achieving an $R^2$ score of 0.972 on DrivAerNet and 0.957 on DrivAerNet++; For surface pressure prediction, TripNet achieves the lowest $\ell_2$ error of 20.25\% on DrivAerNet and 20.05\% on DrivAerNet++. For surface wall shear stress prediction, it also attains the lowest error of 22.07\% on DrivAerNet++ and the second-lowest error of 21.80\% on DrivAerNet.  Table ~\ref{tab:model_comparison} presents a comprehensive comparison between TripNet and previous models. 

    \item TripNets are scalable to different aerodynamic simulation tasks without significant modifications to the model architecture. A U-Net + MLP handles surface and 3D flow fields, while a lightweight CNN predicts drag coefficients. TripNets learn the geometry-flow interactions from the triplanes, enabling accurate flow field predictions while minimizing the computational overhead.


\end{enumerate}

These contributions highlight the effectiveness and scalability of our approach for high-fidelity aerodynamic analysis. To the best of the authors' knowledge, this is the first implementation of the triplane representation to solve partial differential equations (PDEs), as all existing methods have exclusively utilized it for generative modeling.

\begin{figure*}[h!]
    \centering
    \includegraphics[width=\linewidth]{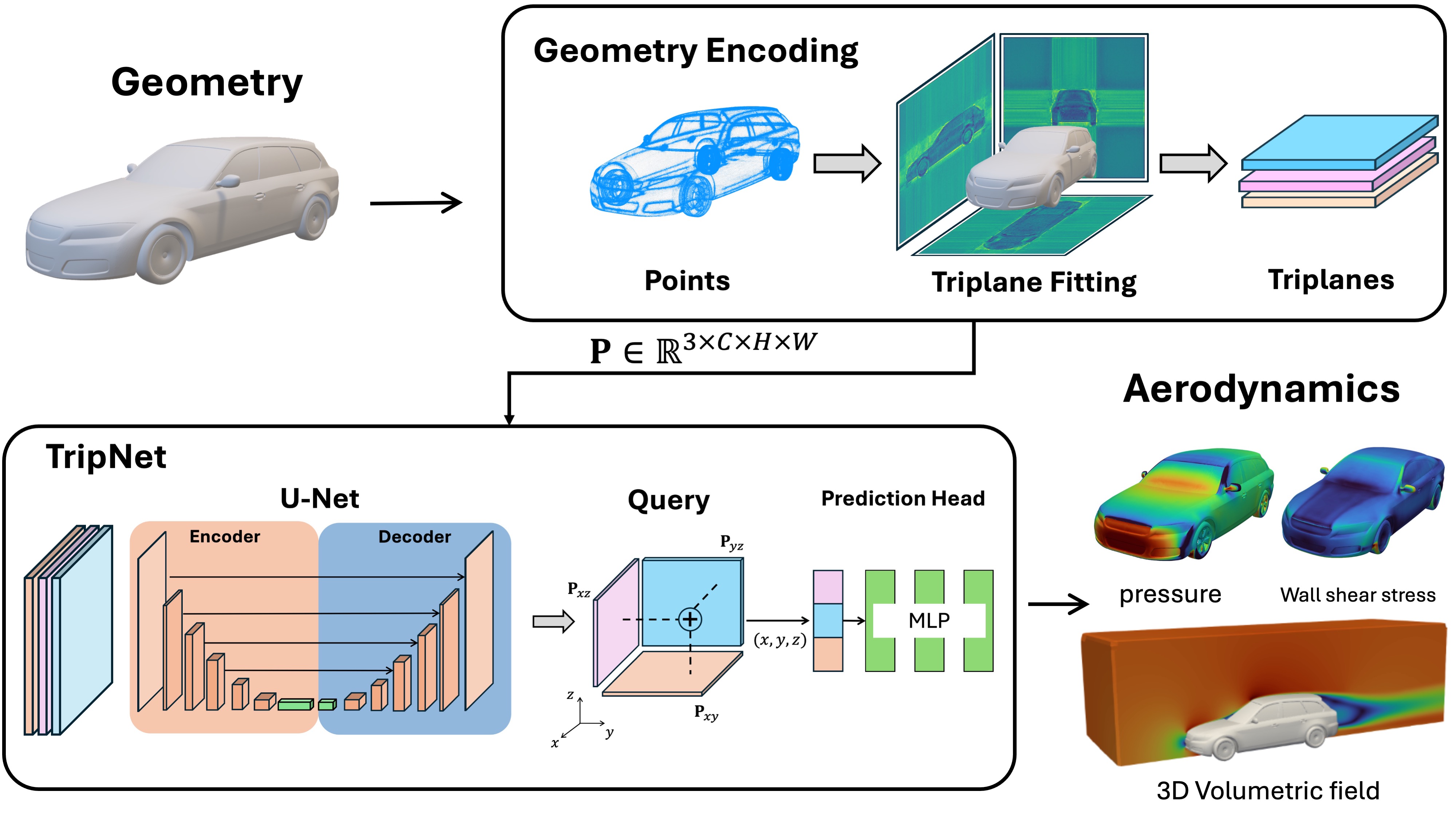}
    \caption{Illustration of Triplane Networks and their role in CFD tasks. The figure demonstrates how triplane features encode spatial information for aerodynamic analysis, showcasing their applications in predicting surface pressure, wall shear stress, and full 3D flow fields.}
    \label{fig:main_figure}
\end{figure*}

        
    

\begin{table*}
    \caption{A comprehensive comparison of various 3D geometric deep learning models for solving Navier-Stokes equations.}
    \centering
    \resizebox{\textwidth}{!}{%
    \begin{tabular}{lcccccccc}
        \toprule
        \multirow{2}{*}{Model} & \multicolumn{3}{c}{CFD Tasks} & \multirow{2}{*}{Geometry} & \multirow{2}{*}{Representation} & \multirow{2}{*}{Prediction}  & \multirow{2}{*}{Prediction Beyond} & \multirow{2}{*}{Memory Cost} \\
        \cmidrule(lr){2-4}
        & $C_d$ & $P$,$\tau_w$ & $U$ & Representation& Type & Fashion & Input Points & Independent on Resolution\\
        \midrule
        ResNeXt~\cite{song2023surrogate} & \bcheck  & \btimes & \btimes& Images & Explicit & Images & \btimes & \btimes \\
        PointNet~\cite{qi2017pointnet} & \bcheck  & \bcheck & \btimes & Point Cloud & Explicit &Input Nodes & \btimes & \btimes \\
        MeshGraphNet~\cite{pfaff2020learning} & \bcheck  & \bcheck & \btimes & Mesh & Explicit & Graph Nodes & \btimes & \btimes \\
        X-MeshGraphNet~\cite{nabian2024x} & \bcheck  & \bcheck & \bcheck & Mesh $+$ Point Cloud & Explicit & Graph Nodes & \btimes & \btimes\\
        Transolver~\cite{wu2024transolver} & \bcheck  & \bcheck & \bcheck & Point Cloud $+$ SDF & Explicit & Input Nodes & \btimes & \btimes \\
        FigConvNet~\cite{choy2025factorized} & \bcheck  & \bcheck & \btimes & Point Cloud & Explicit & Input Nodes & \btimes & \btimes \\
        GINO~\cite{li2024geometry} & \bcheck & \bcheck & \bcheck & Point Cloud $+$ SDF & Explicit & Query & \bcheck & \btimes\\
        TripNet (Ours) & \bcheck  & \bcheck & \bcheck & Triplane & \textbf{Implicit} & Query & \bcheck & \bcheck \\
        \bottomrule
    \end{tabular}
    }
\label{tab:model_comparison}
\end{table*}

\section{Method} 
\label{network} 

In this section, we introduce our methodology. We begin by describing the triplane representation, which serves as the foundation of our approach. We then give details about TripNets for different aerodynamic prediction tasks, including aerodynamic drag prediction, surface field prediction for pressure and wall shear stress (WSS), and volume flow field prediction.

\subsection{Triplane Representation} 
Triplane representation has recently emerged as a powerful approach in 3D representation learning for scene reconstruction~\cite{wu2024blockfusion}, 3D geometry~\cite{zhang2024tar3d} and avatar generation ~\cite{wang2023rodin,sun2023next3d,zhang2025rodinhd} because of its balance between computational efficiency and ability to capture fine-grained details in 3D objects. Three axis-aligned and orthogonal feature planes are used to encode 3D information, offering a compact but expressive way to handle complex 3D shapes and textures. Models such as GANs and diffusion models ~\cite{shue20233d, chan2022efficient} can be trained on these triplane representations to generate 3D neural fields with high diversity and quality. Building on this foundation, the Roll-out diffusion network (Rodin) \cite{wang2023rodin} rolls out the triplanes into a single 2D feature plane and performs 3D-aware diffusion to generate 3D digital avatars. Compared to 1D latent vector, voxel, or grid-based representations, triplane representations offer greater memory efficiency while preserving spatial information, providing a 3D-aware encoding that better captures geometric and structural details.


In this work, we aim to convert the 3D car designs to triplanes for TripNet training. Firstly we build the occupancy field for each 3D object by sampling 5 million points uniformly in the domain and another 5 million points near the surface. Next, we followed the two-stage fitting method NFD~\cite{shue20233d} used to get the triplanes. Each 3D object is represented by triplanes $\mathbf{P}_{xy}, \mathbf{P}_{yz}, \mathbf{P}_{xz}\in \mathbb{R}^{C \times H \times W}$. In the fitting, we project the coordinate  $\mathbf{x} \in \mathbb{R}^3$ onto the triplanes and aggregate the resulting feature vectors to predict the value of occupancy field through an MLP decoder ($MLP_{\Phi}$ with parameters $\Phi$). The occupancy function $occ:R^3 \rightarrow \{0,1\}$ is defined as:
\begin{equation}
    occ(x,y,z) = MLP_{\phi}\{\mathbf{P}_{xy}(x,y) + \mathbf{P}_{yz}(y,z) + \mathbf{P}_{xz}(x,z)\}
\end{equation}

In stage one, the triplane features and MLP decoder are jointly trained on $N=500$ objects. $J=500K$ points are sampled from the 10M points. The loss function is: 
\begin{equation}
\mathcal{L}_{\text{NAIVE}} = \sum_{i}^{N} \sum_{j}^{J} \left\| \text{occ}^{(i)} \left( \bm{x}_j^{(i)} \right) - O_j^{(i)} \right\|_2
\end{equation}

To mitigate potential compression artifacts in triplane-based 3D representations, we draw inspiration from Neural Field Diffusion (NFD)~\cite{shue20233d}, where regularization terms are introduced to simplify the data manifold that the diffusion model must learn. It has been shown that these regularization terms effectively reduce significant artifacts in 3D asset generation. We incorporate the same regularization terms in our method to enhance the quality of the triplanes. For further details, interested are encouraged to refer to the original paper~\cite{shue20233d}. The training objective of the 3D reconstruction task for triplane fitting is as follows: 
\begin{equation}
    \mathcal{L} = \mathcal{L}_{\text{NAIVE}} + \lambda_{1}R_{TV} + \lambda_{2}R_{norm} + \lambda_{3}R_{EDR}
\end{equation}
where the total variation (TV) regularization term, \( R_{TV} \), ensures that the feature planes do not contain spurious high-frequency information. The \( L_2 \)-norm regularization term, \( R_{norm} \), controls the value range of the feature planes within \([-1,1]\). The explicit density regularization (EDR) term, \( R_{EDR} \), promotes the smoothness of the feature planes. In the second stage, the MLP decoder, \( MLP_{\Phi} \), is frozen, and only the triplanes of each car are trained individually using the same loss function.

 Triplanes as a geometric representation, allows for the reconstruction of the original geometry with minimal information loss. By capturing the full geometric details of the object, triplanes enable accurate predictions across multiple aerodynamic tasks. This compact representation operates with a complexity of $\mathcal{O}(n^2)$, making it significantly more efficient and scalable than existing 3D geometry learning models which have a complexity of $\mathcal{O}(n^3)$. Due to computation limitations, existing models often resort to subsampling mesh nodes or points when handling high-resolution data, leading to a loss of geometric details.  As a result, TripNets are not constrained by these limitations and can efficiently serve as a surrogate model for CFD simulations. TripNets utilize a convolutional backbone to process triplane representations efficiently, enabling a unified approach for diverse aerodynamic predictions. This backbone extracts multi-scale geometric features using 2D convolutions, providing a consistent representation for different tasks.  From this shared backbone, task-specific heads extend to different predictions: a lightweight CNN maps the triplane features to a scalar drag coefficient, a U-Net refines features for surface fields (pressure and wall shear stress) and 3D velocity fields as shown in Figure~\ref{fig:main_figure}.

\subsection{Drag Coefficient Task}
To predict the drag coefficient ($C_d$), we employ a lightweight convolutional neural network (CNN) to efficiently process triplane representations\footnote{model architecture can be found in Figure~\ref{fig:Triplane_CNN} in the Appendix}. The input to our model consists of raw triplane features $\mathbf{P}=[\mathbf{P}_{xy}, \mathbf{P}_{yz}, \mathbf{P}_{xz}] \in \mathbb{R}^{3 \times 32 \times 128 \times 128}$. Since each feature plane has a channel depth of 32, stacking the three triplanes and taking it as the input would make the model large and inference slow. We apply channel-wise $max$, $min$, and $mean$ pooling on the raw triplane. Each pooling operator $\Theta:\mathbb{R}^{32 \times 128 \times 128} \rightarrow \mathbb{R}^{3 \times 128 \times 128}$ compresses the channel depth of original triplane $\mathbf{P}_{ij} $ from 32 to 3. Three feature planes $\mathbf{\tilde{P}}_{ij}$ where $ij \in \{xy, yz, xz\}$ are stacked channel-wise for drag coefficient prediction:

\begin{equation}
\begin{aligned}
    \mathbf{\tilde{P}}_{ij} = max(\mathbf{P}_{ij})  \oplus  min(\mathbf{P}_{ij})  \oplus  mean(\mathbf{P}_{ij})
\end{aligned}
\end{equation}

The drag model $f_\Theta:\mathbb{R}^{9 \times 128 \times 128} \rightarrow \mathbb{R}$ takes compressed triplanes as input and predicts the aerodynamic drag coefficient: 
\begin{equation}
    C_d = f_\Theta(\mathbf{\tilde{P}}_{xy} \oplus \mathbf{\tilde{P}}_{yz} \oplus \mathbf{\tilde{P}}_{xz})
\end{equation}

Our model consists of five convolutional layers, each followed by batch normalization and max-pooling layers to progressively extract hierarchical features and reduce spatial dimensions. The extracted features are then flattened and passed through three fully connected layers, transforming the compressed triplane representation into an accurate drag coefficient prediction. This simple yet effective CNN-based architecture allows our model to efficiently capture the aerodynamic properties encoded in triplane features while maintaining computational efficiency. Our approach demonstrates that triplane representation, combined with a lightweight CNN, is sufficient for state-of-the-art drag coefficient prediction, outperforming traditional mesh- and point-based methods in both accuracy and inference speed.


\subsection{Surface Field and Volume Flow Field Tasks}
Predicting surface fields (e.g., pressure and wall shear stresses) and 3D flow fields (e.g., velocity components) is significantly more challenging than predicting a single scalar quantity like the drag coefficient. These tasks require capturing spatially coherent and physically consistent field distributions while maintaining both local precision and global aerodynamic context. To achieve this, we design a hybrid model architecture that combines a U-Net and a Multi-Layer Perceptron (MLP):
\begin{equation}
    \mathbf{V} = F_{\Phi} (\mathbf{P}, x, y, z)
\end{equation}

where $\mathbf{P} = \mathbf{P}_{xy} \oplus  \mathbf{P}_{yz} \oplus  \mathbf{P}_{xz} \in \mathbb{R}^{96  \times 128 \times 128}$. For surface pressure, $\mathbf{V} \in \mathbb{R}$, whereas for surface wall shear stress and 3D velocity field prediction, $\mathbf{V} \in \mathbb{R}^3$.

The U-Net takes as input the concatenated raw triplane features and learns to predict surface and volumetric fields, potentially capturing how these features influence global and local aerodynamic behaviors. By leveraging an encoder-decoder structure with skip connections, the U-Net effectively captures multi-scale dependencies, enabling the model to refine feature representations that are critical for aerodynamic field predictions. In addition to triplane features, the model explicitly incorporates spatial coordinates as an additional input. This allows the network to query the triplane representation at specific locations, extracting localized geometric and aerodynamic information.

The queried triplane features are then concatenated channel-wise and passed into an MLP, which maps these enriched features to predict local aerodynamic quantities, such as pressure ($P$), wall shear stress ($\tau_{x}$, $\tau_{y}$ and $\tau_{z}$), or velocity components ($U_x$, $U_y$ and $U_z$). By combining global feature learning through U-Net and localized feature refinement through MLP, our architecture enables high-fidelity, spatially consistent aerodynamic field predictions. This approach significantly outperforms traditional methods by leveraging the implicit triplane representation, making it more computationally efficient while maintaining superior accuracy across both surface and volumetric field predictions.

\subsection{Specialization Vs Generalization }

In this paper, we evaluate and compare model performance on both DrivAerNet and DrivAerNet++. DrivAerNet serves as a specialization benchmark, focusing on a single car category with limited geometric variation, making it suitable for studying performance in constrained design spaces and class-specific predictions. In contrast, DrivAerNet++ presents a significantly more challenging generalization task, encompassing multiple car categories and a broader diversity of designs. This enables us to assess each model’s ability to generalize across varying shapes and aerodynamic configurations, transitioning from class-specific to cross-class predictions.

\section{Results}
We evaluate our method on DrivAerNet~\cite{elrefaie2024drivaernet} and DrivAerNet++~\cite{elrefaie2024drivaernet++} datasets across aerodynamics tasks including drag coefficient predictions, surface field pressure, and 3D flow field simulations. These datasets were chosen because they are the largest industry-standard datasets available with extensive geometric variations and high-fidelity simualations, making them well-suited for data-driven methods.
The drag model is trained on a single RTX 4090 GPU for 1 hour, while the surface and 3D flow field models are trained on four H100 GPUs over a period of 1.5 days.

\subsection{Aerodynamic Drag Prediction}
We evaluate TripNets for drag coefficient prediction, comparing their performance against existing methods based on point clouds, graphs, and other deep learning architectures, as presented in Table \ref{tab:drivaernet_comparison}. Across both datasets, TripNets consistently achieves the highest accuracy, outperforming previous state-of-the-art models. TripNets establish a new state-of-the-art $R^2$ score of 0.972 on the DrivAerNet dataset and 0.957 on the DrivAerNet++ dataset, demonstrating a strong correlation with ground truth values while maintaining lower error metrics.

The superior performance of TripNets stems from the triplane representation, which preserves full geometric information in an implicit form, unlike traditional mesh, voxel, or point-cloud methods that suffer from downsampling losses. This richer and more expressive feature space enables precise aerodynamic predictions using a lightweight CNN model. While prior approaches attempted to overcome geometric limitations with handcrafted features, such as multi-view image rendering~\cite{song2023surrogate}, these methods introduce human bias and are constrained by predefined viewpoints. In contrast, TripNets autonomously learn task-specific representations, ensuring optimized and highly structured features for aerodynamic prediction, leading to enhanced accuracy and robustness.

\begin{table*}[h!]
\footnotesize
\centering
\caption{Performance comparison on drag coefficient prediction across various models using different data representations, including point clouds, graphs, and our triplane representation. The evaluation is conducted on the DrivAerNet test set, which includes 600 industry-standard car designs. Models are ranked based on $R^2$ performance.}
\label{tab:drivaernet_comparison}
\begin{tabular}{lcccc}
\hline
\multirow{2}{*}{\textbf{Model}} & \textbf{MSE $\downarrow$} & \textbf{MAE $\downarrow$} & \textbf{Max AE $\downarrow$} & \multirow{2}{*}{\textbf{$R^2$ $\uparrow$}}   \\ 
 & $(\times 10^{-5})$ & $(\times 10^{-3})$ & $(\times 10^{-2})$ & \\ \hline
 PointNet~\cite{qi2017pointnet} & 12.0  & 8.85 & 10.18  & 0.826   \\
  GCNN~\cite{kipf2016semi} & 10.7 & 7.17 & 10.97 & 0.874 \\ 
PointNet++~\cite{qi2017pointnet++} & 7.813 & 6.755 & 3.463 & 0.896   \\
RegDGCNN~\cite{elrefaie2024drivaernet} & 8.01 & 6.91  & 8.80  & 0.901   \\
PointBERT~\cite{yu2022point} & 6.334 & 6.204 & 2.767 & 0.915  \\
DeepGCN~\cite{li2019deepgcns} & 6.297 & 6.091 & 3.070 & 0.916   \\
MeshGraphNet~\cite{pfaff2020learning} & 6.0 & 6.08 & 2.965 & 0.917   \\
AssaNet~\cite{qian2021assanet} & 5.433 & 5.81 & 2.39 & 0.927  \\
PointNeXt~\cite{qian2022pointnext}  & 4.577 & 5.2 & 2.41 & 0.939  \\ 
FIGConvNet~\cite{choy2025factorized} & 3.225 & 4.423 & 2.134 & 0.957   \\ 
\hline
\textbf{TripNet (Ours)}  & \underline{2.602} & \underline{4.030} & \underline{1.268} & \underline{0.972} \vspace{1mm} \\ \hline
\end{tabular}
\end{table*}

\subsection{Surface Field}

Beyond drag coefficient prediction, we extend the triplane representation to surface field predictions, such as pressure and wall shear stress, which are crucial for optimizing aerodynamic performance in automotive and aerospace design. Accurate flow field predictions enable faster, more efficient design iterations compared to costly traditional simulations. Our experiments reveal that directly applying the raw triplane features to surface pressure and wall shear stress field prediction introduces significant challenges. The predicted surface fields exhibit noticeable discontinuities and high-frequency noise, particularly in regions of high curvature and complex flow interactions.

\begin{table*}[h!]
    \footnotesize
    \centering
    \setlength{\tabcolsep}{2.5pt} 
\caption{Performance comparison on surface pressure field prediction across various models. The evaluation is conducted on the unseen test set of both the DrivAerNet and DrivAerNet++ dataset, which includes 600 and 1200 industry-standard car designs}    \vspace{2mm}
    \label{tab:surface_filed_pressure}
    \begin{tabular}{llccccc}
    \hline
    \multirow{2}{*}{\textbf{Dataset}} & \multirow{2}{*}{\textbf{Model}} & \textbf{MSE $\downarrow$} & \textbf{MAE$\downarrow$} &\multirow{2}{*}{\textbf{Max AE$\downarrow$}} & \textbf{Rel L2} & \textbf{Rel L1} \\
    & &  $(\times 10^{-2})$  & $(\times 10^{-1})$ & & \textbf{(\%) $\downarrow$}&\textbf{(\%)$\downarrow$}\\
    \hline
    \multirow{4}{*}{DrivAerNet} &  RegDGCNN~\cite{elrefaie2024drivaernet} & 9.01 & 1.58  &  13.09 & 28.49 & 26.29 \\
 & FigConvNet~\cite{choy2025factorized} & 4.38 & 1.13 & 5.73 & 20.98& 
 18.59 \\
  & Transolver~\cite{wu2024transolver} & 5.37 & 1.28 & 5.89  & 22.52& 21.31 \\
    & \textbf{TripNet (Ours)} & \underline{4.23} & \underline{1.11} & \underline{5.52} & \underline{20.35}& \underline{18.52}  \vspace{1mm} \\ 
    \hline  
     \multirow{4}{*}{DrivAerNet++\hspace{10pt}} & 
 RegDGCNN~\cite{elrefaie2024drivaernet} & 8.29 & 1.61 & 10.81 & 27.72& 26.21\\

    & FigConvNet~\cite{choy2025factorized} & \underline{4.99} & \underline{1.22} & 6.55 & 20.86 & 21.12 \\ 
    & Transolver~\cite{wu2024transolver} & 7.15 & 1.41 & 7.12 & 23.87 & 22.57 \\
    & \textbf{TripNet (Ours)} & 5.14 & 1.25 & \underline{6.35} &\underline{20.05} & \underline{20.93}\\
    \hline     
    
    \end{tabular}
\end{table*}

\begin{table*}[ht]
    \footnotesize
    \centering
    \setlength{\tabcolsep}{2.5pt} 
\caption{Performance comparison on surface wall shear stress field prediction across various models on the DrivAerNet and DrivAerNet++ dataset}    
    \label{tab:surface_field_wss}
    \begin{tabular}{l@{\hspace{4mm}}lccccc}
    \toprule
    \multirow{2}{*}{\textbf{Dataset}} & \multirow{2}{*}{\textbf{Model}} & \textbf{MSE $\downarrow$} & \textbf{MAE$\downarrow$} &\multirow{2}{*}{\textbf{Max AE$\downarrow$}} & \textbf{Rel L2} & \textbf{Rel L1} \\
    & &  $(\times 10^{-2})$  & $(\times 10^{-1})$ & & \textbf{(\%) $\downarrow$}&\textbf{(\%)$\downarrow$}\\
    \hline
    \multirow{4}{*}{DrivAerNet} & RegDGCNN~\cite{elrefaie2024drivaernet} & 12.13 & 1.94 &  11.53 & 34.70 & 35.53  \\
 & FigConvNet~\cite{choy2025factorized} & 9.10 & 1.79 & 7.85 & 22.01& 18.59 \\
 & Transolver~\cite{wu2024transolver} & 8.59 & 1.69 & 8.10  & \underline{21.51}& \underline{16.92} \\
    &\textbf{TripNet (Ours)} & \underline{8.14} & \underline{1.66} & \underline{7.47} & 21.80& 17.39 \\
    \hline
    \multirow{4}{*}{DrivAerNet++} & RegDGCNN~\cite{elrefaie2024drivaernet} & 13.82 & 3.64 & 11.54 & 36.42 &36.57\\
 & FigConvNet~\cite{choy2025factorized} & 9.86 & 2.22 & \underline{7.84} & 22.32 & 17.65\ \\
 & Transolver~\cite{wu2024transolver} & \underline{8.95} & \underline{2.06} & 8.45  & 22.49 & 17.65 \\
     & \textbf{TripNet (Ours)} & 9.52 & 2.15 & 8.14 & \underline{22.07}  & \underline{17.18} \\    
    \bottomrule
    \end{tabular}
\end{table*}

\begin{figure*}[h!]
    \centering
    \includegraphics[width=\linewidth]{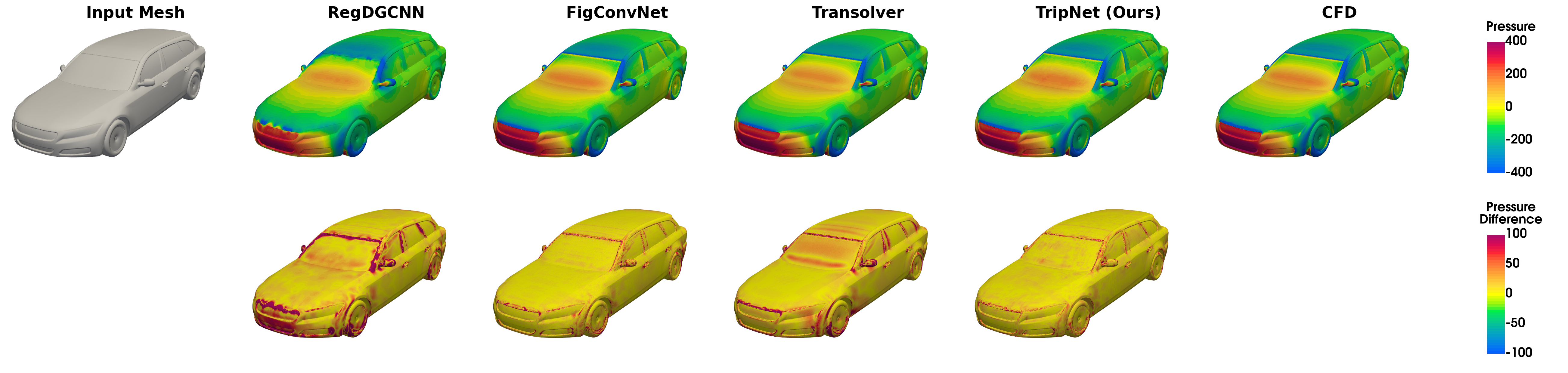}
    \caption{Comparison of pressure predictions and errors for car design E\_S\_WWC\_WM\_094 (estateback with smooth underbody, wheels closed, and with mirrors) from the unseen test set. The first row shows the input mesh followed by predictions from RegDGCNN, FigConvNet, Transolver, TripNet (ours), and the ground-truth CFD. The second row highlights the absolute difference between the predictions and ground-truth CFD, visualizing the absolute error distribution for each model.}
    \label{fig:E_S_WWC_WM_094_isometric}
\end{figure*}

These artifacts arise because triplane features are optimized for volumetric representations rather than structured surface fields. Unlike explicit 3D meshes or point clouds with well-defined geometric mappings, triplanes interpolate from three orthogonal planes, causing projection artifacts on the car’s surface. Moreover, small-scale aerodynamic structures, such as boundary layers and flow separations, require high-resolution features that raw triplane embeddings fail to capture. To address this, we use a U-Net+MLP architecture, where U-Net refines spatial coherence and MLP enables precise query-based field predictions, reducing noise and improving accuracy.

Figure~\ref{fig:E_S_WWC_WM_094_isometric} illustrates the pressure predictions and error distributions for the car design E\_S\_WWC\_WM\_094 from the unseen test set, demonstrating the accuracy of different models, including our TripNet, in comparison to ground-truth CFD results. 
TripNet demonstrates state-of-the-art performance in both wall shear stress and surface pressure prediction tasks across the DrivAerNet and DrivAerNet++ datasets. As shown in Table~\ref{tab:surface_filed_pressure} and Table~\ref{tab:surface_field_wss}, TripNet achieves the lowest or second-lowest values across all key evaluation metrics—including MSE, MAE, Max AE, $\ell_1$ and $\ell_2$ error compared to established baseline models such as RegDGCNN, FigConvNet, and Transolver. These results indicate that TripNet matches or exceeds the performance of current state-of-the-art approaches, highlighting its effectiveness in high-fidelity aerodynamic field prediction tasks.

\begin{figure*}[h!]
    \centering
    \includegraphics[width=\linewidth]{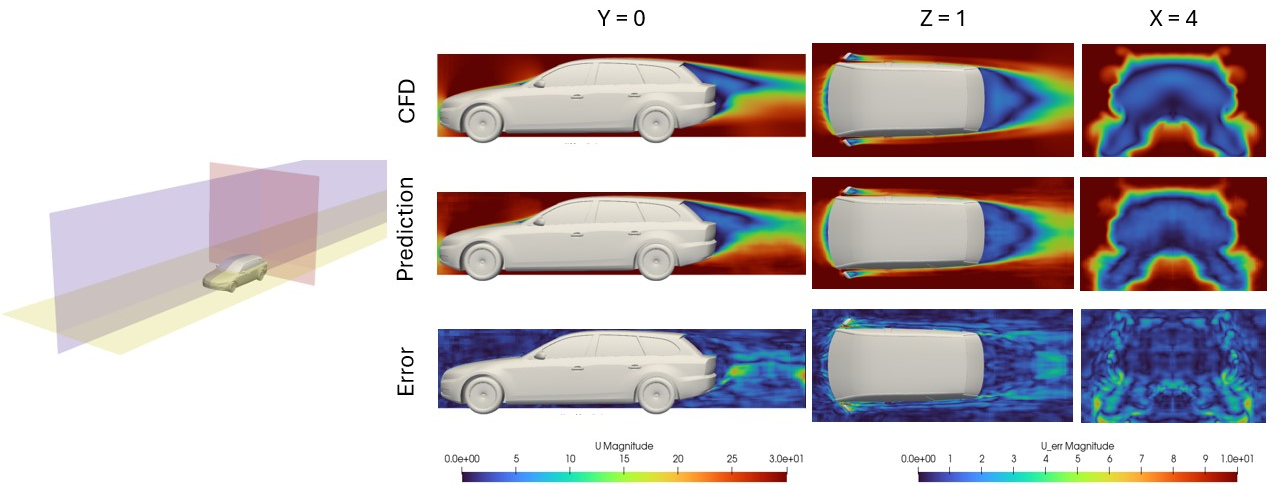}
\caption{Comparison of 3D flow field predictions, including velocity magnitude ($U$) and errors, for the car design E\_S\_WWC\_WM\_094 (estateback with smooth underbody, with wheels closed, and with mirrors) from the unseen test set. The visualization includes three planes: $y=0$ (symmetry plane),  $z=1$, and $x=4$ (wake of the car). The first row shows the ground truth CFD results, the second row displays our model's predictions, and the final row illustrates the absolute error.}
    \label{fig:3d_field}
\end{figure*}

\subsection{Volume Flow Field}
In addition to predicting surface-related aerodynamic properties such as surface pressure and wall shear stresses\footnote{Surface field predictions on wall shear stress are presented in Appendix.~\ref{sec:additional_results} due to page limitation.}, our method extends the use of triplane features to predict the full 3D flow field around a car. Traditional CFD methods require solving complex partial differential equations on volumetric meshes, making full 3D flow field prediction computationally expensive and time-consuming. Our approach leverages the implicit nature of triplane features, enabling efficient data-driven flow field estimation while maintaining geometric fidelity.

\begin{table}[h!]
\caption{Error metrics for 3D flow field prediction, comparing our model's predictions with the ground-truth CFD. The table includes the velocity components in the $x$, $y$, and $z$ directions, as well as the total velocity magnitude.}
\footnotesize
\centering
\setlength{\tabcolsep}{2pt} 
\renewcommand{\arraystretch}{1.2} 
\begin{tabular}{lccccc}
\hline
\multirow{2}{*}{\textbf{Field}} & \multirow{2}{*}{\textbf{MSE $\downarrow$}} & \multirow{2}{*}{\textbf{MAE $\downarrow$}} & \multirow{2}{*}{\textbf{Max AE $\downarrow$}} & \textbf{Rel L1} & \textbf{Rel L2} \\
&  &  &  & \textbf{(\%)$\downarrow$} & \textbf{(\%)$\downarrow$} \\
\hline
$\mathbf{U_x}$ & 6.69 & 1.49 & 31.31 & 7.15 & 10.71 \\
$\mathbf{U_y}$ & 2.24 & 0.86 & 25.06 & 28.97 & 35.34 \\
$\mathbf{U_z}$ & 2.39 & 0.86 & 22.07 & 31.12 & 36.39 \\
$\mathbf{U}$ & 6.71 & 1.52 & 28.09 & 6.88 & 10.39 \\
\hline
\end{tabular}
\label{tab:flow_field}
\end{table}

While triplane features are initially extracted to encode surface geometry information, they inherently capture global geometric context, including regions corresponding to empty space around the car. When revisiting the geometry reconstruction task used to fit the triplane, we observe that the triplane features not only represent surface properties but also provide valuable information about the spatial structure of the surrounding flow field. This insight allows us to extend triplane processing beyond the car surface, leveraging its information-rich representation to predict full 3D velocity fields near the car. Table~\ref{tab:flow_field} shows the performance metrics on  volumetric flow field predictions observed between TripNet predictions and the simulation, averaged over the test set of DrivAerNet++. TripNets achieve an $\ell_1$ error of 7.15\% and $\ell_2$ error of 10.7\% for $U_x$ prediction. A higher $\ell_1$ and $\ell_2$ error is observed for $U_y$ and $U_z$ due to their sparse distribution and lower magnitude compared to $U_x$. The training, validation, and test dataset for volumetric flow field prediction consists of 8,000 simulations, amounting to approximately 16TB of high-fidelity CFD data.

Figure~\ref{fig:3d_field} presents a comparison of 3D flow field predictions for velocity magnitude ($U$) in three cross-sectional planes of the car design E\_S\_WWC\_WM\_094. The figure visualizes the ground truth CFD results, our model's predictions, and the corresponding absolute error across the symmetry plane ($y=0$), the wake region ($x=4$), and a vertical slice cutting the car body ($z=1$). Our model accurately predicts the wake flow in the rear of the car as well as the wake generated by the side mirrors, demonstrating its effectiveness in capturing complex aerodynamic flow structures. Additional comparison between TripNet and CFD results is provided in Appendix. ~\ref{appendix_3d_flow}.

\begin{figure}[h]
    \centering
    \begin{minipage}{0.325\textwidth}
        \centering
        \includegraphics[width=\textwidth]{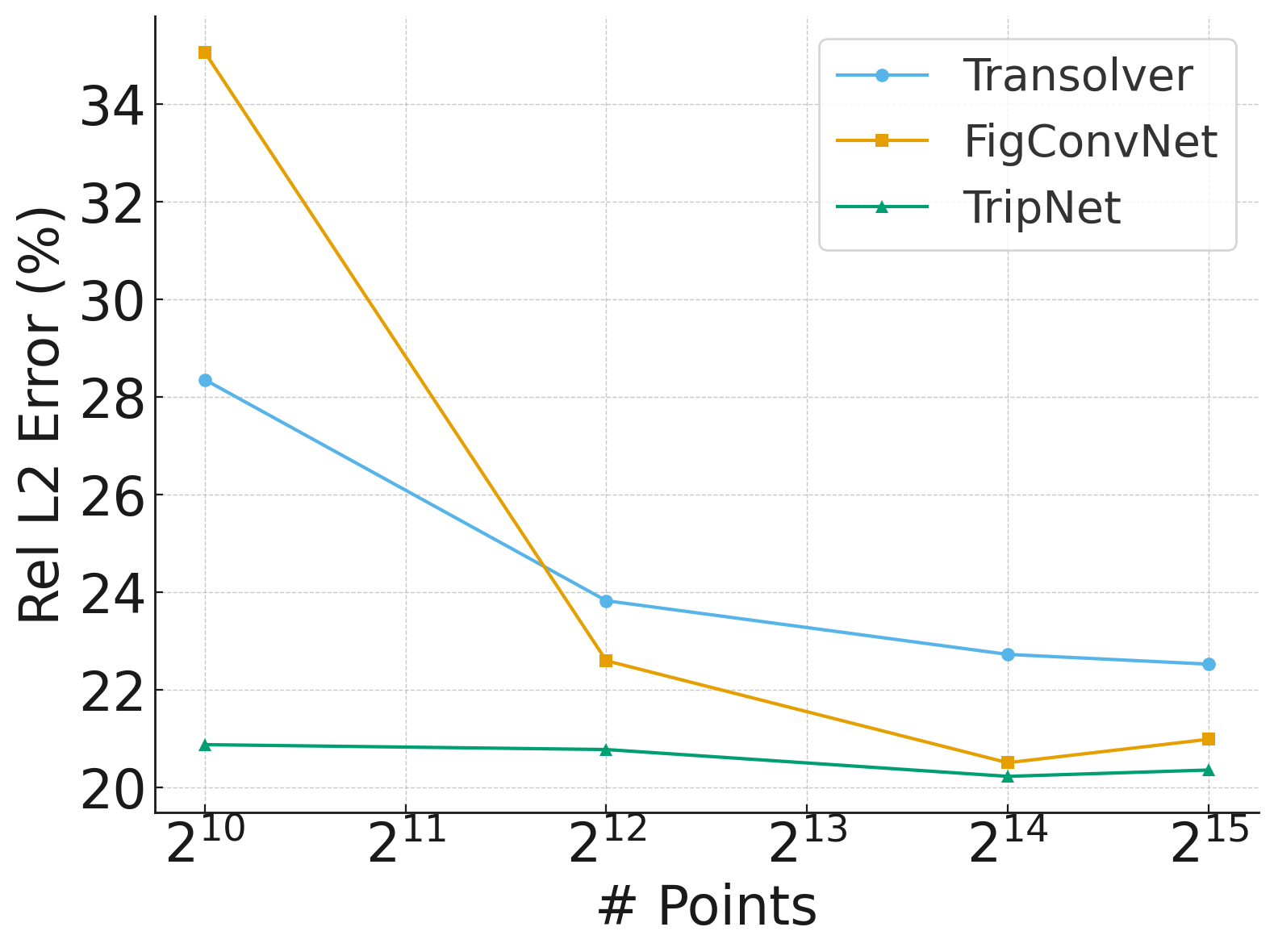}
        \caption*{(a)}
        \label{fig:rel_l2}
    \end{minipage}
    \hfill
    \begin{minipage}{0.325\textwidth}
        \centering
        \includegraphics[width=\textwidth]{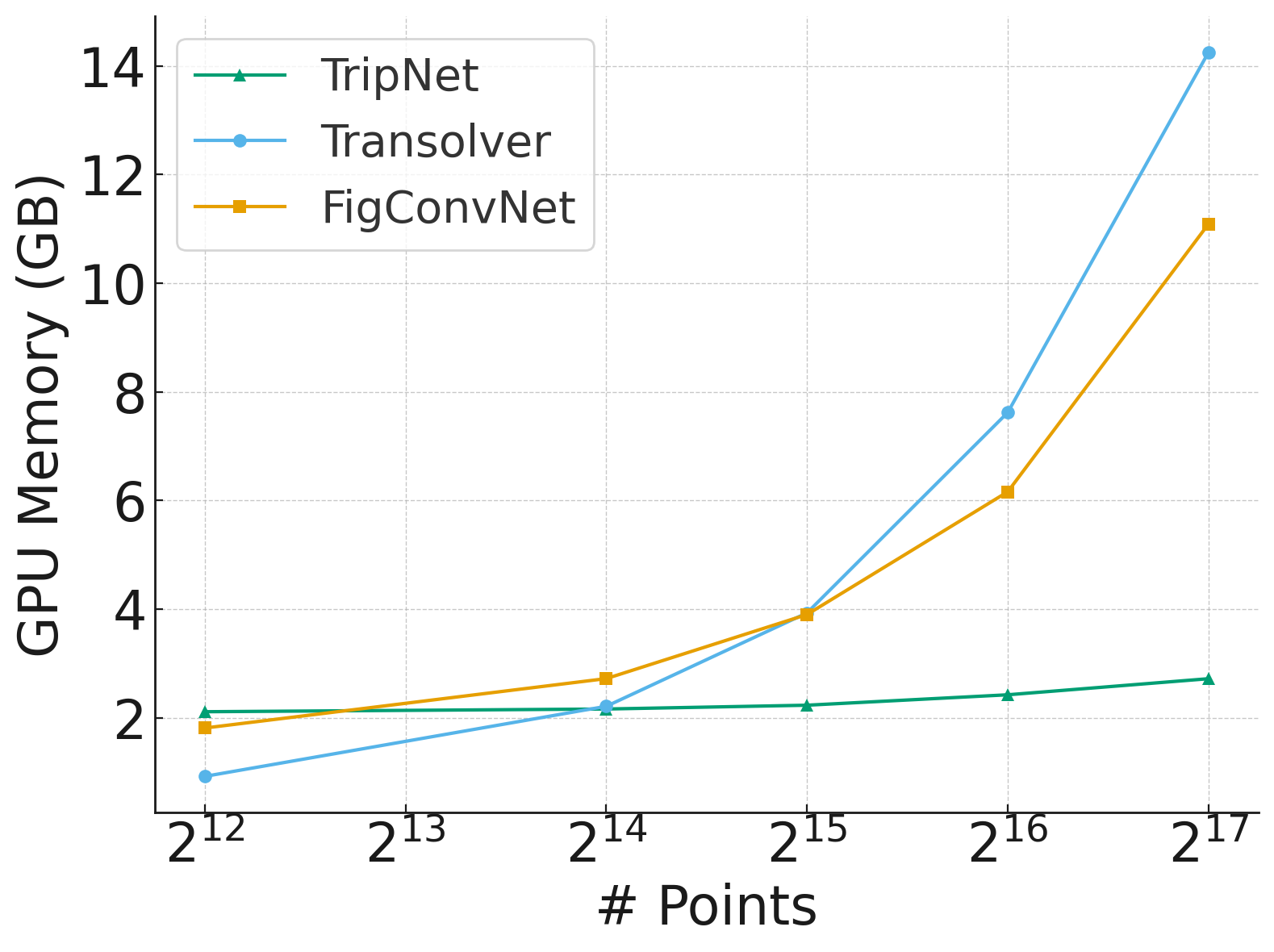}
        \caption*{(b)}
        \label{fig:gpu_memory}
    \end{minipage}
    \hfill
    \begin{minipage}{0.325\textwidth}
        \centering
        \includegraphics[width=\textwidth]{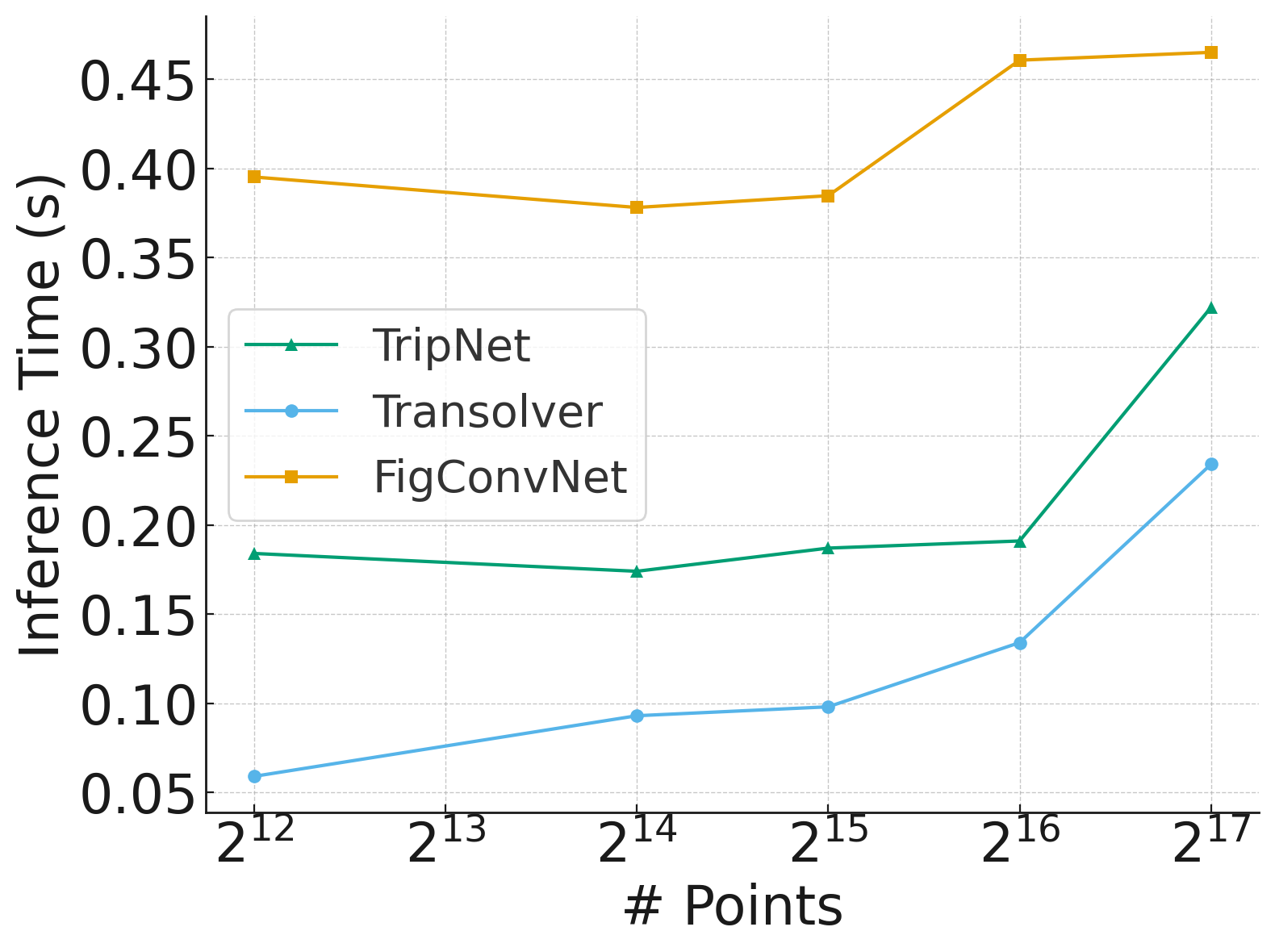}
        \caption*{(c)}
        \label{fig:inference_time}
    \end{minipage}
    \caption{Scalability comparison of TripNet, FigConvNet, and Transolver across different point resolutions. (a) Relative L2 Error vs. Number of Points. (b) GPU Memory Usage vs. Number of Points; (c) Inference Time vs. Number of Points}
    \label{fig:scalability_all}
\end{figure}


\section{Conclusion}  
\label{conclusion}  

In this work, we introduced TripNets as surrogate models for aerodynamic simulations, including drag coefficient prediction, surface pressure and wall shear stress (WSS) estimation, and 3D flow field prediction. Leveraging triplane features—a compact, high-dimensional, and implicit representation widely used in 3D generation tasks—we demonstrated their effectiveness in aerodynamic modeling, achieving state-of-the-art performance.  

Our approach efficiently encodes aerodynamic properties into triplane features, enabling multiple predictive tasks. For drag coefficient prediction, we utilize compressed triplane features obtained through channel-wise max, min, and average pooling, feeding them into a lightweight CNN-based model. On the DrivAerNet dataset, our model achieves an $R^2$ score of 0.97, requiring only one hour of training and 0.01 seconds for inference per car. 


TripNet outperforms or matches state-of-the-art models on both pressure and wall shear stress prediction tasks across the DrivAerNet and DrivAerNet++ datasets. For pressure prediction, TripNet achieves the lowest relative $\ell_2$ errors of  $20.35\%$ (DrivAerNet) and $20.05\%$  (DrivAerNet++), compared to $20.98\%$ and $20.86\%$ for FigConvNet, and $22.52\%$  and $23.87\%$  for Transolver. On wall shear stress prediciton, TripNet also leads with relative $\ell_2$ errors of 22.07\%, while FigConvNet reports 22.32\%, and Transolver lags behind at 22.49\% on DrivAerNet++ dataset. As shown in Figure ~\ref{fig:scalability_all}, TripNet offers significantly improved computational efficiency and scalability: TripNet uses \textbf{4.1$\times$ less GPU memory} and is \textbf{1.4$\times$ faster} in inference compared to FigConvNet at the highest resolution (131{,}072 points). Unlike FigConvNet and Transolver, whose GPU memory usage grows with input resolution, TripNet maintains nearly constant memory usage and inference time owing to its resolution-independent triplane representation. In addition, for volumetric flow field prediction, the structured and dense triplane feature representation used by TripNet is inherently more suitable than the sparse point cloud representation employed by FigConvNet, enabling more accurate and complete predictions across the full 3D domain.

Furthermore, we demonstrate that triplane features encode not only geometric information but also localized flow characteristics, enabling accurate volumetric flow field predictions near the car surface. These results highlight the potential of TripNets as a scalable and efficient alternative to traditional CFD solvers for aerodynamic analysis.

\section{Limitation and Future Work}
\label{limitation}
Despite its strong performance, our method has several limitations. The triplane fitting is performed separately for each geometry, so the model’s ability to generalize to topologically distinct shapes (e.g., F1 cars) or different flow regimes beyond the training Reynolds number remains unverified. Additionally, the model does not explicitly enforce physical constraints such as mass conservation or turbulence-model consistency, which may lead to non-physical artifacts, particularly in regions with sparse training data. Future work will explore incorporating physics-informed constraints and training on more diverse geometries and flow conditions to improve generalization and ensure physically consistent predictions.

\bibliographystyle{plainnat}
\bibliography{neurips_2025}


\newpage
\appendix

\renewcommand{\thefigure}{\arabic{figure}}
\renewcommand{\thetable}{\arabic{table}}
\setcounter{figure}{0}
\setcounter{table}{0}

\section*{Technical Appendices and Supplementary Material}

\startcontents[appendices]  
\phantomsection

\section*{Table of Contents}
\printcontents[appendices]{}{1}{}  

\newpage

\section{Related Work}
\label{related work}
In this section, we introduce Reynolds-Averaged Navier-Stokes (RANS) equations for engineering design simulations, review deep learning models for simulation, discuss the limitations of graph- and point cloud-based neural networks, and finally explore implicit and neural representations. 

\paragraph{RANS for Initial Design Stages}  
Reynolds-Averaged Navier-Stokes (RANS) simulations are a fundamental approach in industrial fluid dynamics, particularly valuable in the early and conceptual design stages due to their balance of accuracy and computational efficiency. Recent datasets for aerodynamic analysis have employed RANS simulations, showcasing a wide range of dataset sizes and mesh resolutions. These include 500 designs in Usama et al.~\cite{Usama2021}, 889 in Umetani et al.~\cite{Umetani2018}, 1,000 in Gunpinar et al.~\cite{Gunpinar2019}, and 1,400 in Remelli et al.~\cite{remelli2020meshsdf}. Larger datasets include 2,474 designs in Song et al.~\cite{song2023surrogate} and 1,121 in Trinh et al.~\cite{trinh20243d}, with mesh resolutions ranging from 130k in Gunpinar et al.~\cite{Gunpinar2019} to 3M in Song et al.~\cite{song2023surrogate}.  The most recent DrivAerNet~\cite{elrefaie2024drivaernet} and DrivAerNet++~\cite{elrefaie2024drivaernet++} datasets were published to address the trade-off between large dataset sizes required for training deep learning models and the simulation fidelity necessary for industrial applications. These datasets comprise 8,000 designs with high-resolution meshes, providing an unparalleled level of detail for evaluating machine learning-based CFD approaches. This combination ensures robust performance benchmarking and facilitates the development of more accurate predictive models.

\paragraph{Deep Learning Models for Simulation:} 
 Song et al \cite{song2023surrogate} adopts a new 3D representation using stacked depth and normal renderings. Six single views, front, rear, top, bottom, right and left, are generated and integrated into a single image to capture the geometric features comprehensively. CNN-based and transformer-based models are used to learn the features from the 2D representation and predict drag coefficients. Jacob et al.~\cite{jacob2021deep} introduced a method using Signed Distance Fields (SDF) in combination with a U-Net archicture to simultaneously predict both the drag and velocity fields. This representation captures surface proximity and enables spatially-aware predictions across the flow field.

Graph Neural Networks (GNNs) have emerged as one of the most popular approaches for learning-based CFD simulations due to their ability to handle irregular geometries and unstructured meshes. However, several challenges remain in their application to large-scale, high-fidelity aerodynamic datasets.  MeshGraphNets (MGNs) have been extensively studied for their generalization capabilities in fluid dynamics, particularly for unseen geometries~\cite{schmocker2024generalization}. A comparative analysis of computational performance on both CPU and GPU systems demonstrated a 6.5x speedup in inference on a CPU and a 100.8x speedup on a GPU, significantly reducing training and inference times for replicating full CFD simulations. However, this study primarily focuses on fluid simulations for standard geometries, utilizing meshes with $\mathcal{O}(10^3)$ nodes. Two key limitations of this approach are (1) the use of relatively simple geometries (e.g., cylinders) and (2) the restricted number of nodes, again at $\mathcal{O}(10^3)$. To address these scalability limitations, X-MeshGraphNet~\cite{nabian2024xmeshgraphnet} presents a multi-scale, mesh-free extension of MGNs. It partitions large graphs and incorporates halo regions for inter-partition message passing. This design supports distributed training with equivalent full-graph performance. X-MeshGraphNet also constructs graphs directly from STL geometry files by generating surface or volumetric point clouds, connected via k-nearest neighbors. It further builds multi-resolution graphs to support long-range interactions, enabling real-time simulation without requiring mesh generation at inference. Despite the complexity, it maintains the predictive accuracy of full-graph GNNs.

Transformer-based solvers such as Transolver~\cite{wu2024transolver} enable PDE resolution on general geometries, operating on meshes from $2^{10}$ to $2^{15}$ nodes. Although they outperform linear transformer variants, scaling to industrial-scale datasets like DrivAerNet++, which contains millions of nodes, remains a challenge.

Neural operator is a family of machine learning models designed the approximate the solutions of PDEs. Unlike neural networks that map inputs to outputs, neural operators are trained to approximate the mapping between function spaces. Among the popular architectures, the Fourier Neural Operator (FNO) ~\cite{li2020fourier} utilizes the Fourier transform to represent the kernel integration and solves PDEs in the spectral domain, achieving computational efficiency on regular grids by leveraging Fast Fourier Transforms (FFT). The Graph Neural Operator (GNO) ~\cite{li2024geometry} adopts a graph representation for kernel integration, making it highly suitable for handling complex geometries and irregular domains.

To summarize the capabilities of these models, Table.~\ref{tab:model_comparison} in the main paper provides a comparative overview of representative 3D deep learning architectures for solving Navier-Stokes equations. The table outlines geometry representations, prediction fashion, and scalability features. TripNet distinguishes itself by using a triplane-based implicit representation that supports query-based predictions beyond the input points and maintains memory efficiency across resolutions.

\paragraph{Implicit and Neural Representations:}
Implicit and neural representations~\cite{serrano2023infinity, catalani2024neural} offer a continuous alternative to traditional discrete approaches in CFD simulations. Unlike GNNs and point cloud-based models, which provide solutions only at discrete mesh nodes, implicit representations enable querying the solution at any point in space (and time, if included as an input feature). This flexibility allows for higher-resolution predictions without being constrained by the mesh structure, making them particularly advantageous for modeling complex aerodynamic flows. de Vito et al.~\cite{de2024implicit} presents an implicit neural representation framework for accurate 3D CFD flow field prediction, leveraging a coordinate-based MLP (backbone-net) to provide a discretization-agnostic, infinite-resolution representation of flow fields for domains with arbitrary topology. Combined with a hyper-net that maps blade surface geometry to the backbone-net's parameters, this approach offers a memory-efficient, highly accurate proxy to CFD solvers, demonstrating strong potential for industrial applications like wing and turbine blade design.


\section{Models Description}
The TripNet is trained on 4 H100 GPUs with 200 epochs. We use a batch size of 32, and for each surface field predictions, we sample 200,000 points in the training. All the other baseline models are trained on one A100 GPU. For a fair comparison, we also trained TripNet on one A100 GPU with a batch size of 8 and the 200,000 sampled points. The training setup is summarized in Table ~\ref{tab:training_setup}.

\begin{table}[h!]
    \footnotesize
    \centering
    \setlength{\tabcolsep}{1.7pt} 
    \caption{Training setup of TripNet and baseline models on one A100 GPU}
    \vspace{2mm}
    \label{tab:training_setup}
    \begin{tabular}{lccccc}
    \hline
    \textbf{Model} & \textbf{parameters} & \textbf{Epochs} & \textbf{Optimizer} & \textbf{Batch size} & \textbf{learning rate} \\
    \hline
    RegDGCNN& 5,750,820 & 150 & Adam & 12 &  1e-3 \\
    FigConvNet & 77,649,570 & 200 &  Adam & 1 &  1e-3\\
    Transolver & 3,860,033 & 200 & Adam & 1 &  1e-3 \\
    TripNet & 24,122,672 & 200 & Adam & 32 &  1e-3 \\

    \hline
    \end{tabular}
\end{table}

\subsection{TripNet Overview}
\label{{sec:TripNet_overview}}

\subsubsection{Triplane Fitting}
\label{{sec:triplane_fitting}}

\begin{figure*}[h!]
    \centering
    \includegraphics[width=\linewidth]{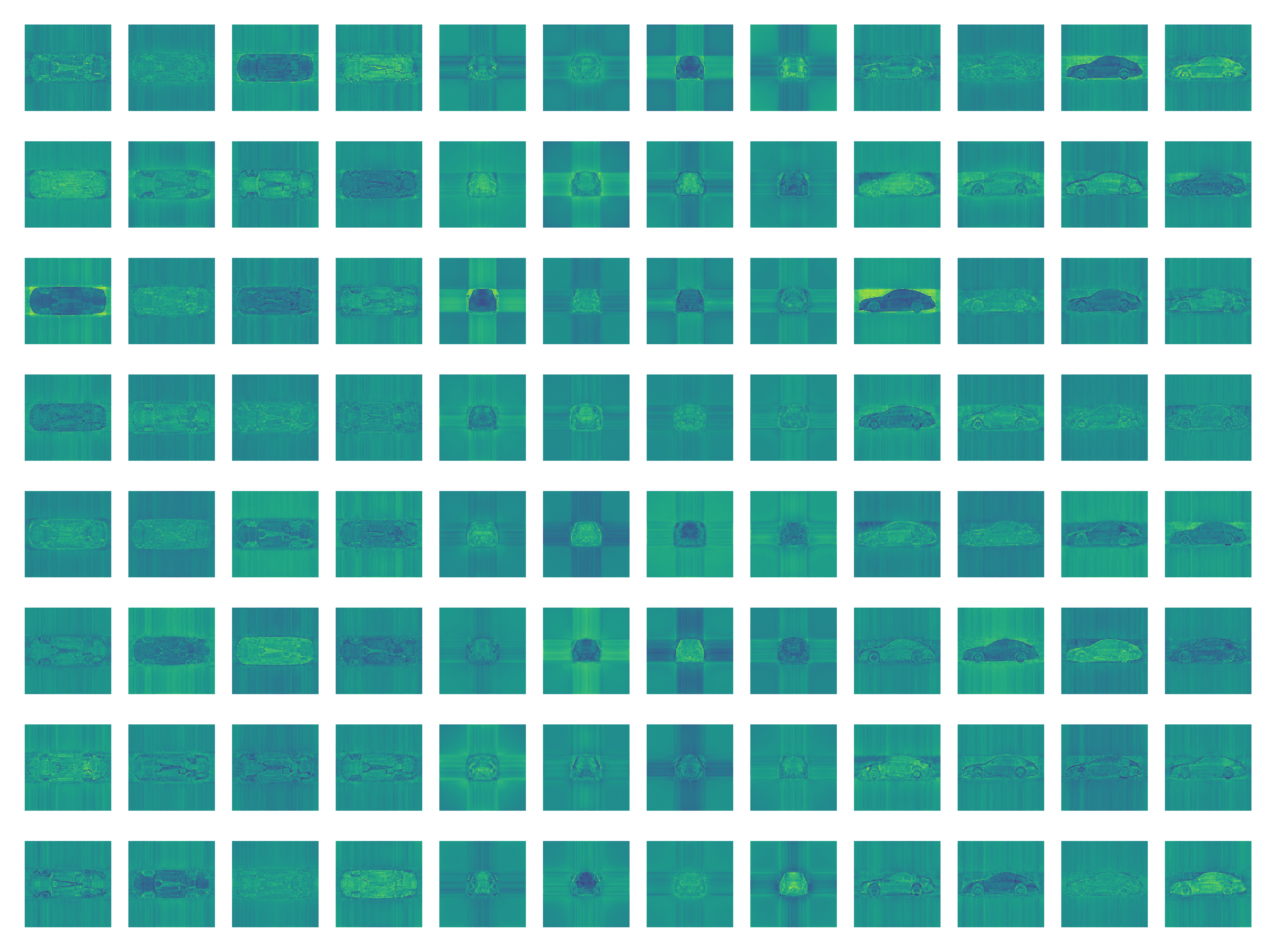}
    \vspace{-5mm}
\caption{Visulization of the raw triplane features of F\_D\_WM\_WW\_1282 (fastback with detailed underbody, with wheels, and with mirrors) used for surface field and 3D flow field predictions.}
    \label{fig:A_vis_triplane}
\end{figure*}

We followed the NDF paper to fit the triplane and adopted a two-stage training to fit the triplanes for cars in the DrivAerNet++ dataset.  In Stage 1, we randomly picked 500 cars and jointly trained the triplane of each car and the shared MLP decoder. We train the first stage with 500 epochs and a learning rate of 1e-3. 
In Stage 2, the MLP decoder is shared and we fit the triplane of each object individually across multiple machines for the sake of efficiency. The fit of each takes 20 seconds to converge with 70 epochs. The triplane obtained from geometry reconstruction task and used for aerodynamics tasks is shown in Figure \ref{fig:A_vis_triplane}.

\subsubsection{Triplane Resolution}
\label{sec:triplane_resolution}

\begin{figure}[h!]
    \centering
    \includegraphics[width=0.8\linewidth]{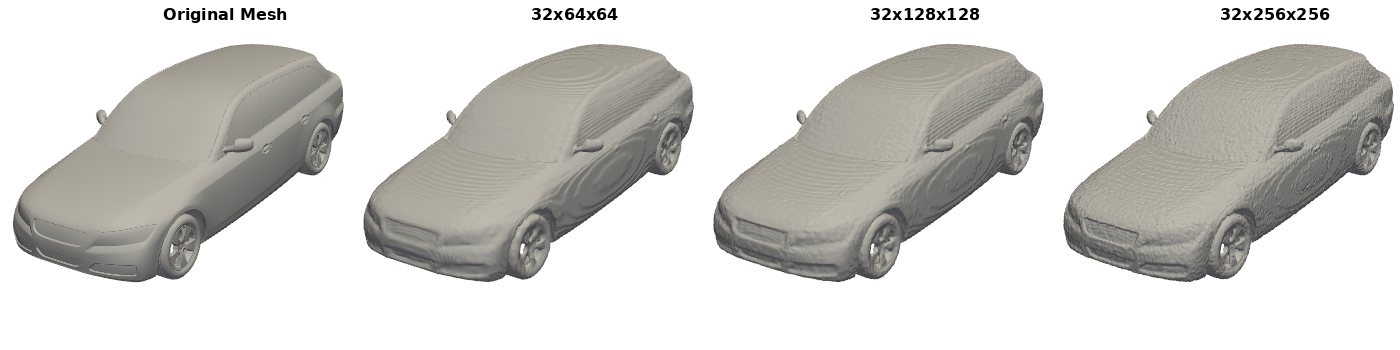}
    \caption{3D Geometry reconstructed by triplanes with different dimensions.}
    \label{fig:append_dim_triplane}
\end{figure}

To evaluate the impact of triplane resolution on 3D geometry reconstruction, we compare different triplane resolutions and analyze their effects on reconstruction quality. Figure ~\ref{fig:append_dim_triplane} illustrates the reconstructed 3D geometries at various triplane resolutions, while Table ~\ref{tab:A_triplane_resolution} provides quantitative metrics, including precision, recall, and F1-score, to assess reconstruction fidelity. Our results show that increasing triplane resolution improves reconstruction accuracy. At a lower resolution of $32 \times 64  \times 64$, the reconstructed geometry retains the overall shape but lacks fine surface details. As the resolution increases to $32 \times 128  \times 128$,  the details become more refined, and the precision, recall, and F1-score improve. At $32\times 256 \times 256$, the reconstruction achieves near-perfect fidelity, closely matching the original mesh, with an F1-score of 0.9936.

We compared the performance of model trained by triplanes with different dimensions and noticed that higher triplane resolutions come with increased memory and computational costs. As summarized in Table ~\ref{tab:Performance_Triplane}, A resolution of \( 32 \times 128 \times 128 \) offers a good balance between reconstruction accuracy and efficiency. These results highlight the trade-off between resolution and computational efficiency, which should be carefully considered when selecting the appropriate triplane representation for downstream tasks.

\begin{table}[ht]
    \footnotesize
    \centering
\caption{Impact of triplane resolution on geometry reconstruction quality. Higher triplane resolutions lead to improved precision, recall, and F1 score, indicating more accurate and detailed geometry representation.}
    \label{tab:A_triplane_resolution}
    \vspace{2mm}
    \begin{tabular}{lccc}
    \hline
     & $32 \times64\times 64$& $32 \times 128\times 128$ & $32 \times 256\times 256$ \\   
     \hline
     Precision & 0.9893 & 0.9914 & 0.9941 \\
     Recall & 0.9870 & 0.9896 & 0.9931 \\ 
     F1 Score & 0.9881 & 0.9905 & 0.9936 \\
     \hline
    \end{tabular}
\end{table}

\begin{table}[h!]
    \footnotesize
    \centering
    \setlength{\tabcolsep}{1.7pt} 
    \caption{Performance of the Triplane network using different triplane resolutions on surface pressure field task on DrivAerNet dataset.}
    \vspace{2mm}
    \label{tab:Performance_Triplane}
    \begin{tabular}{lccccc}
    \hline
    \multirow{2}{*}{\textbf{Resolution}} & \textbf{MSE $\downarrow$} & \textbf{Mean AE$\downarrow$} &\textbf{Max AE$\downarrow$} & \textbf{Rel L2} & \textbf{Rel L1} \\
    &  $(\times 10^{-2})$  & $(\times 10^{-1})$ & & \textbf{Error (\%) $\downarrow$}&\textbf{Error (\%)$\downarrow$}\\
    \hline
    $32 \times 64 \times 64$ & 5.79 & 1.26 & 6.70 & 22.73 & 20.93  \\
    $32 \times 128 \times 128 $ &4.23 & 1.11 & 5.52 & 20.35 & 18.52  \\
    $32 \times 256 \times 256 $ &4.29 & 1.16 & 5.24 & 20.06 & 18.36  \\
    \hline   
    
    \end{tabular}
\end{table}

\subsection{Model Performance}
\label{sec:model_performance}
The scalability of TripNet comes from the query-based prediction. All the baseline models need to build a complete graph to represent the geometry. However, the resolution and point number is limited by the GPU memory when it comes to meshes with millions of nodes in our experiment. TripNet is more flexible because all the geometry is represented by fixed dimensions of triplane feature maps. When the simulation resolution becomes higher, we use bilinear sampling to query features, which could still guarantee the accuracy while avoiding memory issues because the geometry representation cost is independent of mesh and simulation resolution.

\begin{table}[h]
\footnotesize
    \centering
    \caption{Scalability analysis on TripNet, FigConvNet, and Transolver.}
    \label{tab:my_label}
    \begin{tabular}{ccccccc}
    \hline
    \multirow{2}{*}{\textbf{Model}} & \multirow{2}{*}{\textbf{Point number}} & \textbf{MSE $\downarrow$} & \textbf{Mean AE$\downarrow$} &\textbf{Max AE$\downarrow$} & \textbf{Rel L2} & \textbf{Rel L1} \\
    & &  $(\times 10^{-2})$  & $(\times 10^{-1})$ & & \textbf{Error (\%) $\downarrow$}&\textbf{Error (\%)$\downarrow$}\\
    \hline   
        \multirow{4}{*}{Transolver} 
         & 1024 & 7.72 & 1.56 & 3.00  & 28.35 & 26.01\\   
         & 4096 & 5.60 & 1.35 & 3.44  & 23.82 & 22.54 \\
         & 16384  & 5.58 & 1.29 & 5.03  & 22.72 & 21.49\\
         & 32768 & 5.37 & 1.28 & 5.89  & 22.52& 21.31\\
        \midrule
        \multirow{4}{*}{FigConvNet} 
         & 1024 & 11.84 & 2.00 & 3.47 & 35.06 & 33.36\\  
         & 4096  & 5.66 & 1.29 & 3.33 & 22.59 & 21.37\\
         & 16384  & 4.77 & 1.14 & 4.88 & 20.50 & 18.89\\
         & 32768 & 4.38 & 1.13 & 5.73 & 20.98 & 18.59\\
        \midrule
        \multirow{4}{*}{TripNet} 
         & 1024 & 4.11 & 1.15& 3.23  & 20.87& 18.67\\   
         & 4096 &  4.18&  1.17 & 3.41 & 20.77 & 18.93\\
         & 16384  & 4.08 & 1.13 & 5.45 & 20.22 & 18.43\\
         & 32768 & 4.23 & 1.11 & 5.52  & 20.35 & 18.52\\

        \bottomrule
    \end{tabular}
\end{table}

\newpage
\section{Additional Results}
\label{sec:additional_results}

\subsection{Drag coefficient}
\begin{figure*}[h!]
    \centering
    \includegraphics[width=\linewidth]{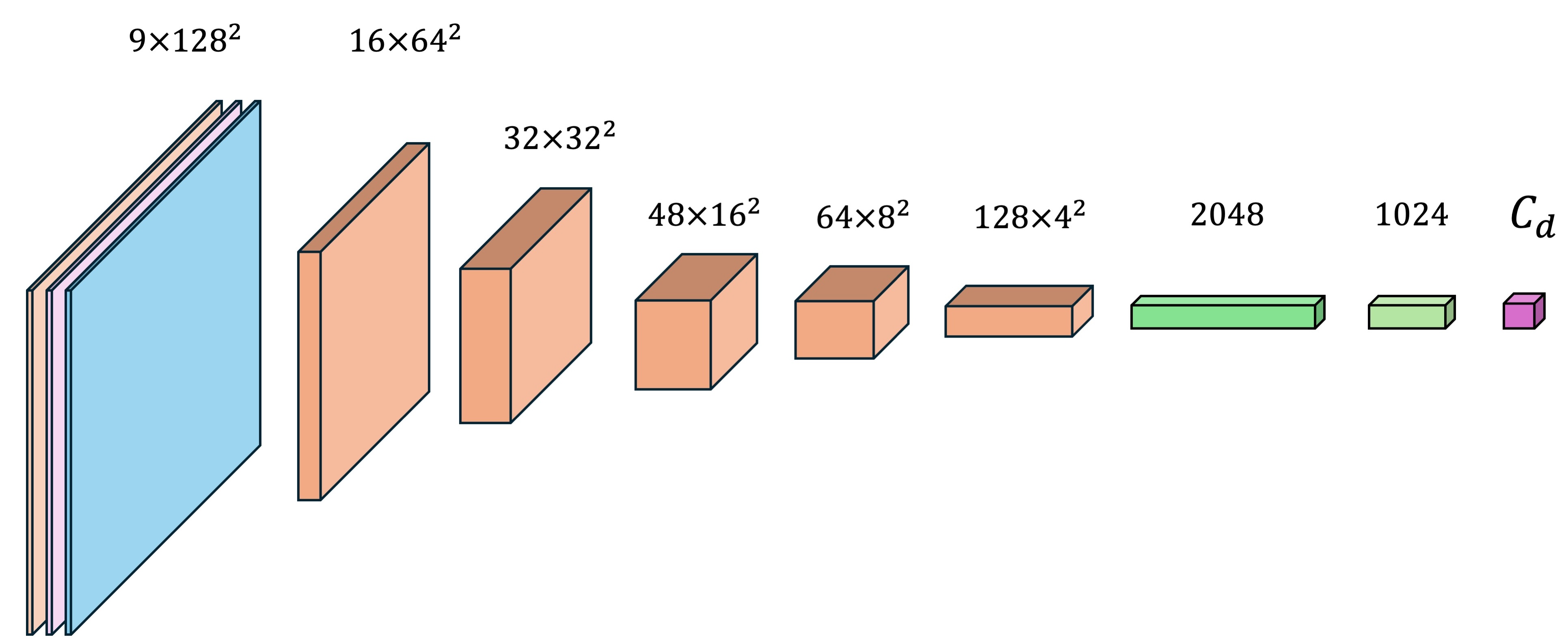}
    \vspace{-5mm}
\caption{Lightweight CNN model for drag coefficient prediction}
    \label{fig:Triplane_CNN}
\end{figure*}

 Figure~\ref{fig:Triplane_CNN} illustrates the architecture of a convolutional neural network designed to predict the drag coefficient ($C_d$) from stacked triplanes of 3D geometry. This surrogate model follows an encoder-style structure, progressively reducing spatial dimensions while increasing feature depth, followed by fully connected layers for regression. The model begins with an input tensor of shape $9 \times 128^2$, representing 9-channel triplanes $\mathbf{\tilde{P}}_{xy}$, $\mathbf{\tilde{P}}_{yz}$ and $\mathbf{\tilde{P}}_{xz}$.  The architecture consists of convolution encoder and flattening dense layers. The output layer predicts a scalar drag coefficient $C_d$. Table ~\ref{tab:drivaernet_plus_plus_drag_comparison} summarizes the performance of different models on drag coefficient prediction on the DrivAerNet++ dataset. TripNet achieves the state-of-the-art performance with $R^2$ score of 0.957.

\begin{table*}[h]
\footnotesize
\centering
\caption{Performance comparison on drag coefficient prediction DrivAerNet++ test set. }

\vspace{2mm}
\label{tab:drivaernet_plus_plus_drag_comparison}
\begin{tabular}{llcccc}
\hline
\multirow{2}{*}{\textbf{Dataset}} & \multirow{2}{*}{\textbf{Model}} & \textbf{MSE $\downarrow$} & \textbf{MAE $\downarrow$} & \textbf{Max AE $\downarrow$} & \multirow{2}{*}{\textbf{$R^2$ $\uparrow$}}   \\ 
 & & $(\times 10^{-5})$ & $(\times 10^{-3})$ & $(\times 10^{-2})$ & \\ \hline

 \multirow{4}{*}{DrivAerNet++}  &GCNN~\cite{kipf2016semi} \textsuperscript{*}& 17.1 & 10.43 & 15.03 & 0.596    \\
&RegDGCNN~\cite{elrefaie2024drivaernet}\textsuperscript{*} & 14.2 & 9.31 & 12.79 & 0.641   \\
&PointNet~\cite{qi2017pointnet}\textsuperscript{*} & 14.9 & 9.60 & 12.45 & 0.643    \\
&\textbf{TripNet (Ours)}   & \underline{9.1} & \underline{7.17} & \underline{7.70} & \underline{0.957} \\ \hline
\end{tabular}
\end{table*}

\newpage


    

\subsection{Pressure Field}

\begin{figure}[h!]
    \centering

    \begin{subfigure}
        \centering
        \includegraphics[width=\linewidth]{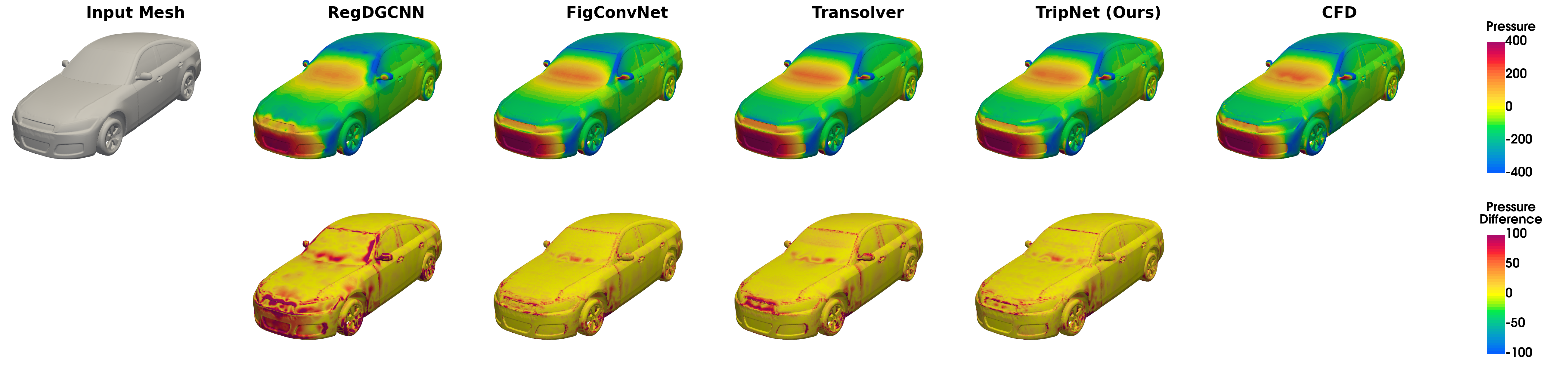}
        \vspace{-4mm}
        \parbox[t]{\textwidth}{\centering (a) Design F\_D\_WM\_WW\_1769 (fastback, detailed underbody, with wheels, and with mirrors).}
        \label{fig:fastback_detailed}
    \end{subfigure}
    \vspace{3mm}
    \begin{subfigure}
        \centering
                \vspace{0.5mm}
        \includegraphics[width=\linewidth]{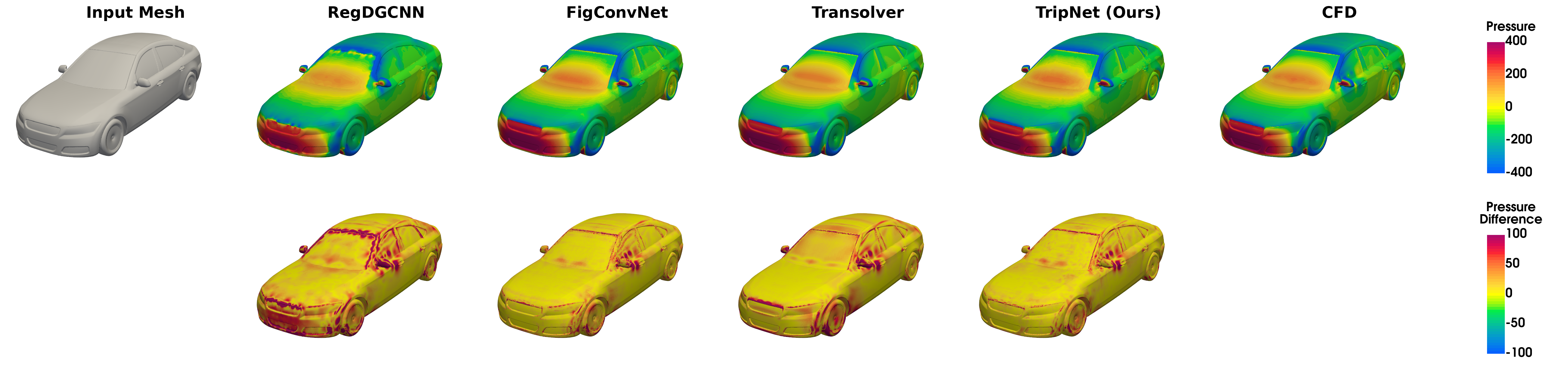}
        \vspace{-4mm}
        \parbox[t]{\textwidth}{\centering (b) Design F\_S\_WWC\_WM\_504 (fastback, smooth underbody, wheels closed, and with mirrors).}
        \label{fig:fastback_smooth}
    \end{subfigure}
    \vspace{3mm}
    \begin{subfigure}
        \centering
                \vspace{0.5mm}

        \includegraphics[width=\linewidth]{Imgs/E_S_WWC_WM_094_isometric.png}
        \vspace{-4mm}
        \parbox[t]{\textwidth}{\centering (c) Design E\_S\_WWC\_WM\_094 (estateback, smooth underbody, wheels closed, and with mirrors).}
        \label{fig:estateback_smooth}
    \end{subfigure}
    \vspace{2mm}
            \begin{subfigure}
        \centering
                        \vspace{0.5mm}
        \includegraphics[width=\linewidth]{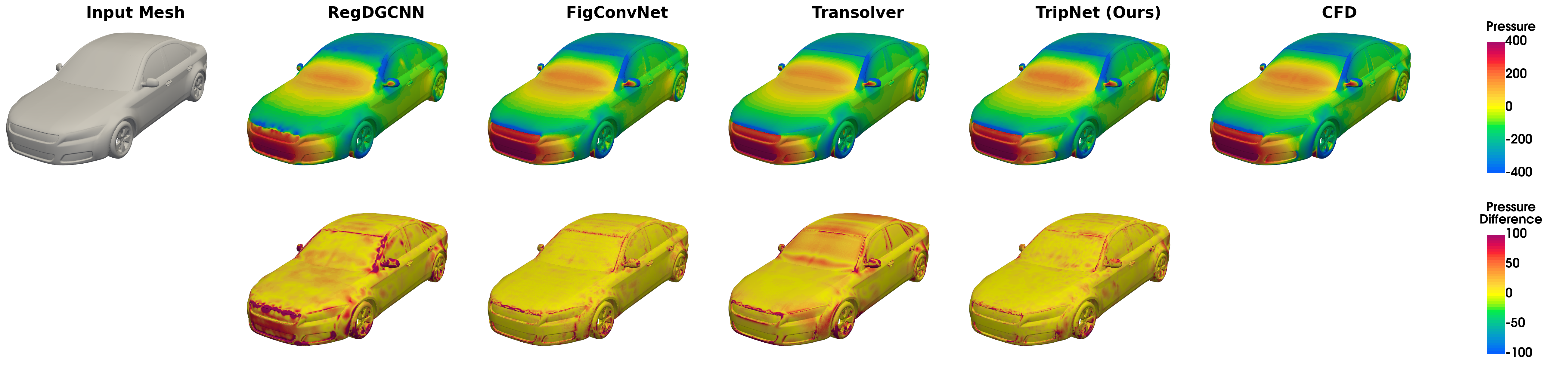}
        \vspace{-4mm}
        \parbox[t]{\textwidth}{\centering (d) Design N\_S\_WW\_WM\_025 (notchback, smooth underbody, with wheels, and with mirrors).}
        \label{fig:notchback_smooth}
    \end{subfigure}
    \vspace{3mm}
    \caption{Comparison of pressure predictions and errors for three car designs from the unseen test set. For each car, the first row shows the input mesh followed by predictions from RegDGCNN, FigConvNet, Transolver, TripNet (ours), and the ground-truth CFD. The second row highlights the absolute difference between the predictions and the ground-truth CFD, visualizing the absolute error distribution for each model.}
    \label{fig:comparison_all}
\end{figure}

\begin{figure}[h!]
    \centering
    \begin{subfigure}
        \centering
        \includegraphics[width=\textwidth]{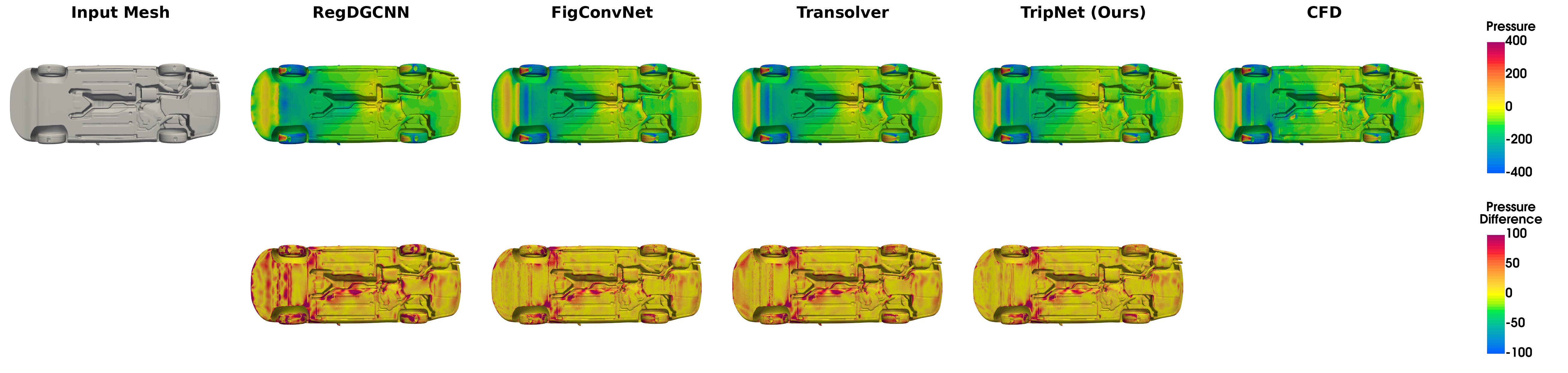}
        \vspace{-4mm}
        \parbox[t]{\textwidth}{\centering (a) Design F\_D\_WM\_WW\_1769 (fastback, detailed underbody, with wheels, and with mirrors).}
        \label{fig:fastback_detailed_underbody}
    \end{subfigure}
    \vspace{3mm}
    \begin{subfigure}
        \centering
        \vspace{0.5mm}
        \includegraphics[width=\textwidth]{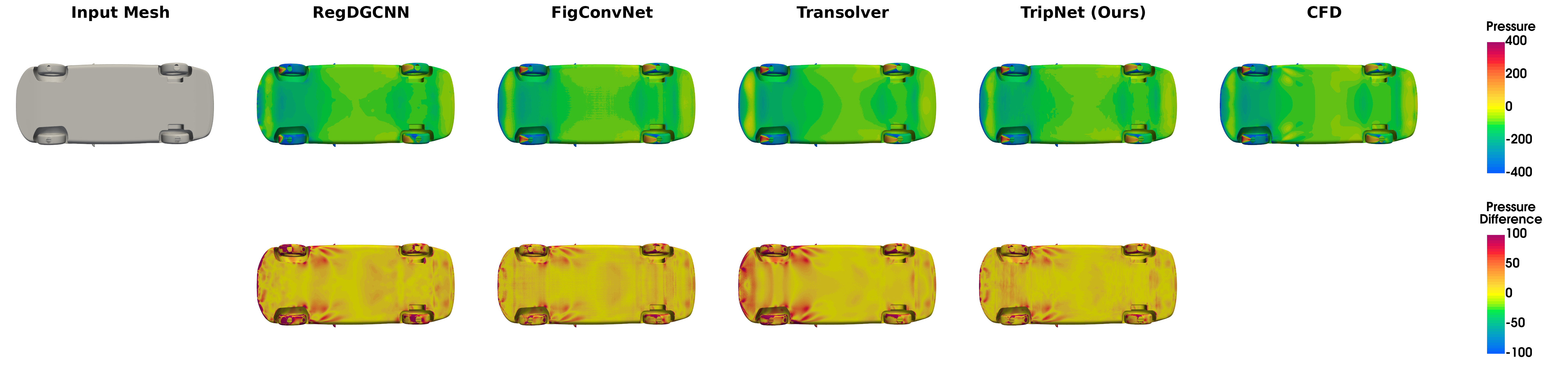}
        \vspace{-4mm}
        \parbox[t]{\textwidth}{\centering (b) Design F\_S\_WWC\_WM\_504 (fastback, smooth underbody, wheels closed, and with mirrors).}
        \label{fig:fastback_smooth_underbody}
    \end{subfigure}
    \vspace{3mm}
    \begin{subfigure}
        \centering
                \vspace{0.5mm}

        \includegraphics[width=\textwidth]{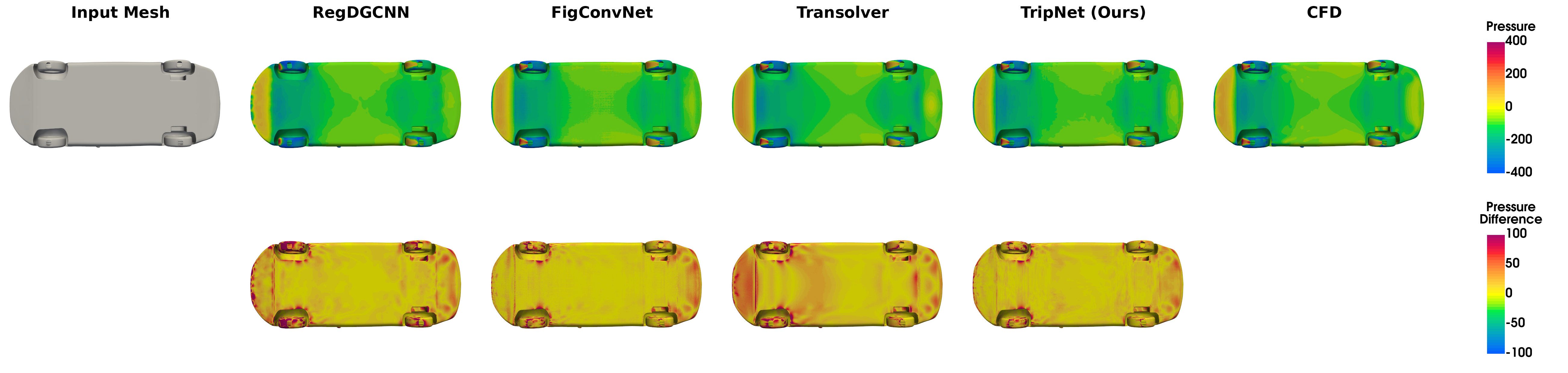}
        \vspace{-4mm}
        \parbox[t]{\textwidth}{\centering (c) Design E\_S\_WWC\_WM\_094 (estateback, smooth underbody, wheels closed, and with mirrors).}
        \label{fig:estateback_underbody}
    \end{subfigure}
    \vspace{2mm}
        \begin{subfigure}
        \centering
                \vspace{0.5mm}
        \includegraphics[width=\linewidth]{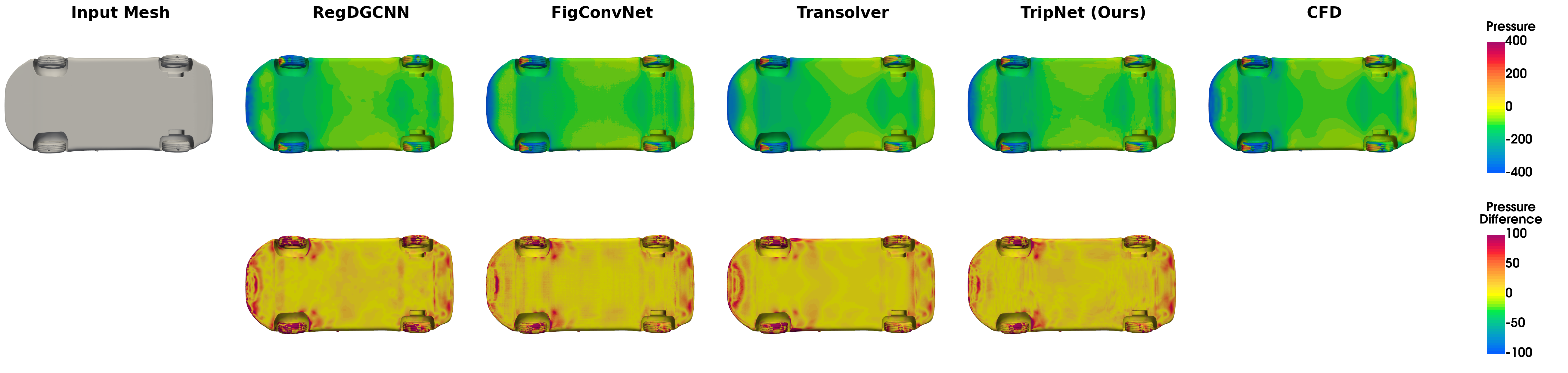}
        \vspace{-4mm}
        \parbox[t]{\textwidth}{\centering (d) Design N\_S\_WW\_WM\_025 (notchback, smooth underbody, with wheels, and with mirrors).}
        \label{fig:notchback_underbody}
    \end{subfigure}
    \vspace{3mm}
    
    \caption{Comparison of pressure predictions and errors for three car designs from the unseen test set. For each car, the first row shows the input mesh followed by predictions from RegDGCNN, FigConvNet, TripNet (ours), and the ground-truth CFD. The second row highlights the absolute difference between the predictions and the ground-truth CFD, visualizing the absolute error distribution for each model.}
    \label{fig:comparison_three_underbody}
\end{figure}

Figures~\ref{fig:comparison_all} and~\ref{fig:comparison_three_underbody} provide a detailed comparison of pressure predictions and errors across multiple deep learning models for three car designs selected from the unseen test set. These figures highlight differences between model predictions and the ground truth CFD results, visualizing the absolute error distribution for each method. We consider various car categories and configurations, including different underbody and wheel setups, to assess the generalization capabilities of deep learning-based aerodynamic prediction methods across diverse design variations.

\clearpage

\subsection{Wall Shear Stress Field}

\begin{figure}[h!]
    \centering
    \begin{subfigure}
        \centering
        \includegraphics[width=\textwidth]{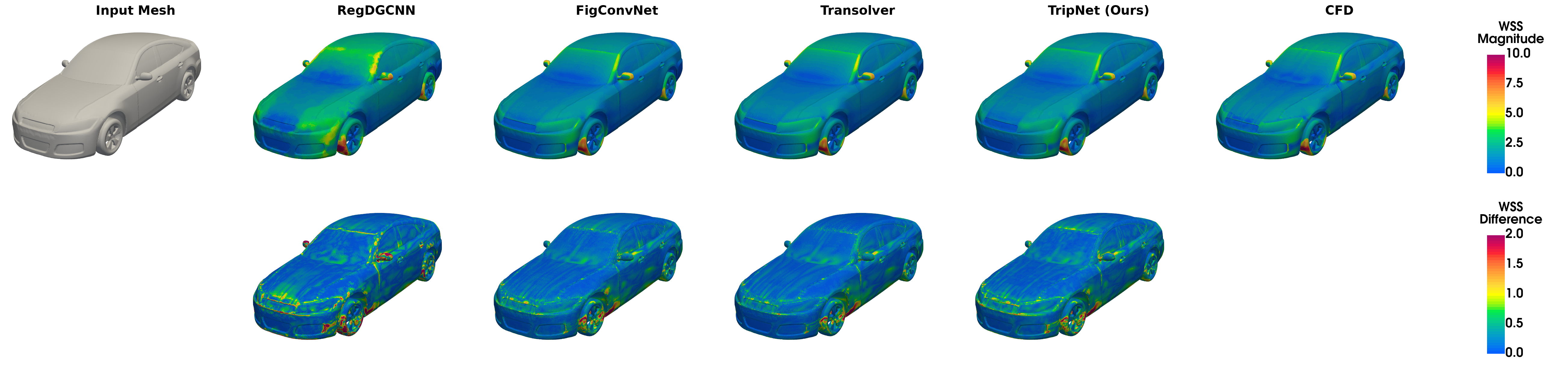}
                \vspace{-4mm}
        \parbox[t]{\textwidth}{\centering (a) Design F\_D\_WM\_WW\_1769 (fastback, detailed underbody, with wheels, and with mirrors).}
        \label{fig:fastback_detailed_isometric_wss}
    \end{subfigure}
    \vspace{3mm}
    \begin{subfigure}
        \centering
        \vspace{0.5mm}
        \includegraphics[width=\textwidth]{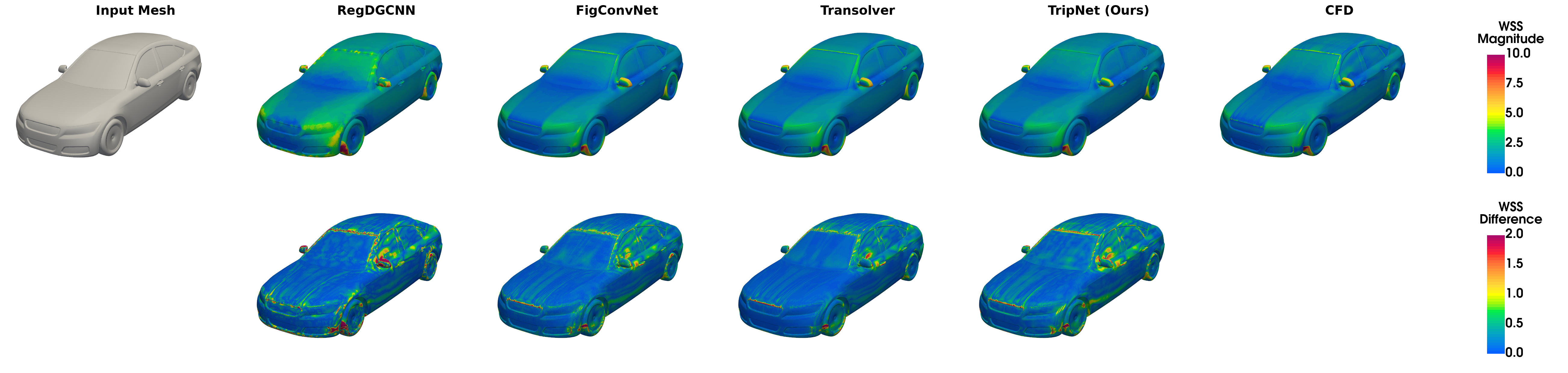}
        \vspace{-4mm}
        \parbox[t]{\textwidth}{\centering (b) Design F\_S\_WWC\_WM\_504 (fastback, smooth underbody, wheels closed, and with mirrors).}
        \label{fig:fastback_smooth_isometric_wss}
    \end{subfigure}
    \vspace{3mm}
    \begin{subfigure}
        \centering
                \vspace{0.5mm}

        \includegraphics[width=\textwidth]{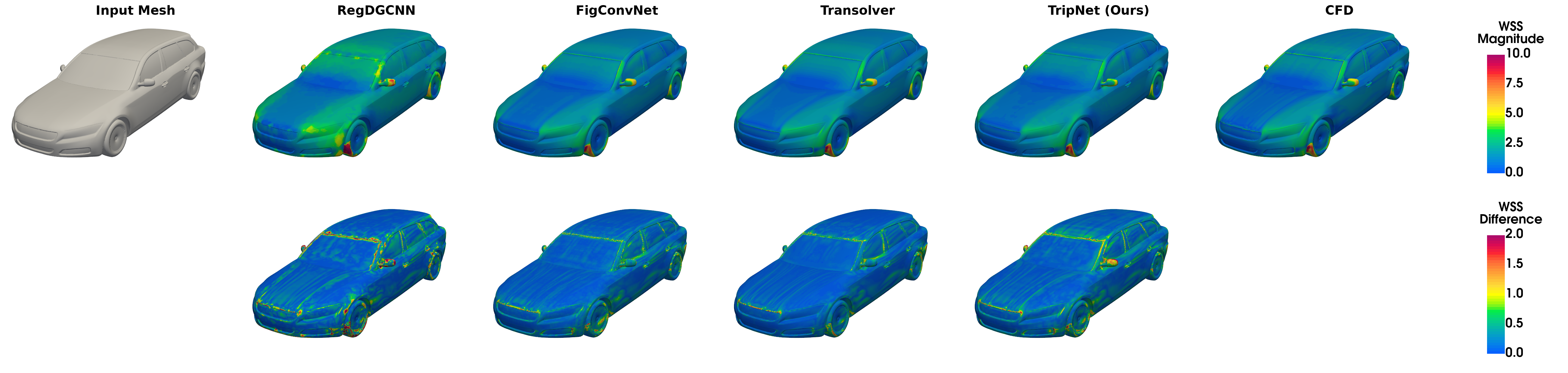}
        \vspace{-4mm}
        \parbox[t]{\textwidth}{\centering (c) Design E\_S\_WWC\_WM\_094 (estateback, smooth underbody, wheels closed, and with mirrors).}
        \label{fig:estateback_isometric_wss}
    \end{subfigure}
    \vspace{2mm}
        \begin{subfigure}
        \centering
                \vspace{0.5mm}

        \includegraphics[width=\textwidth]{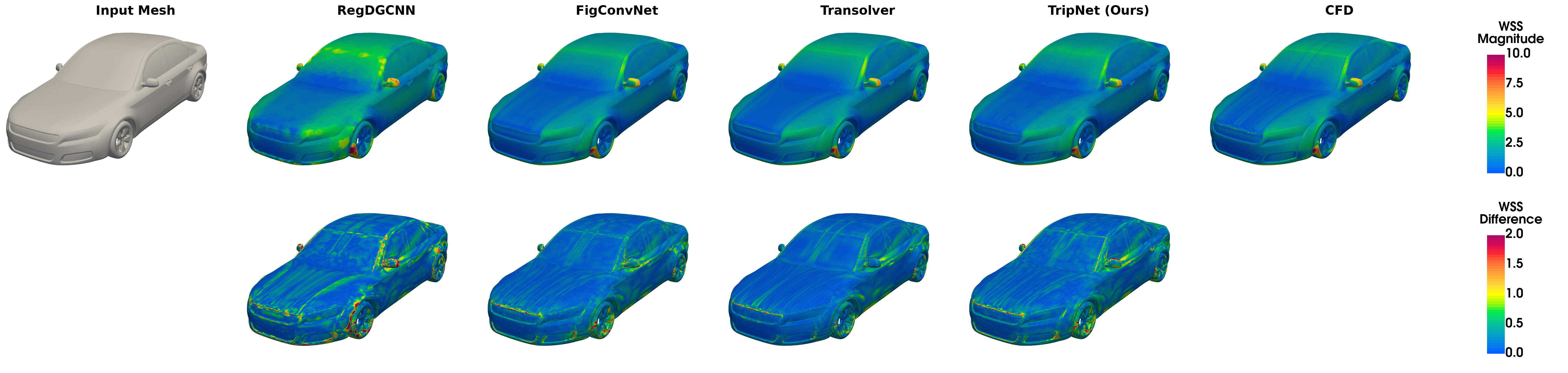}
        \vspace{-4mm}
        \parbox[t]{\textwidth}{\centering (d) Design N\_S\_WW\_WM\_025 (notchback, smooth underbody, with wheels, and with mirrors).}
        \label{fig:notchback_isometric_wss}
    \end{subfigure}
    \vspace{2mm}
    \caption{Comparison of wall shear stress predictions and errors for three car designs from the unseen test set. For each car, the first row shows the input mesh followed by predictions from RegDGCNN, FigConvNet, Transolver, TripNet (ours), and the ground-truth CFD. The second row highlights the absolute difference between the predictions and the ground-truth CFD, visualizing the absolute error distribution for each model.}
    \label{fig:comparison_three_isometric_wss}
\end{figure}

\begin{figure}[h!]
    \centering
    \begin{subfigure}
        \centering
        \includegraphics[width=\textwidth]{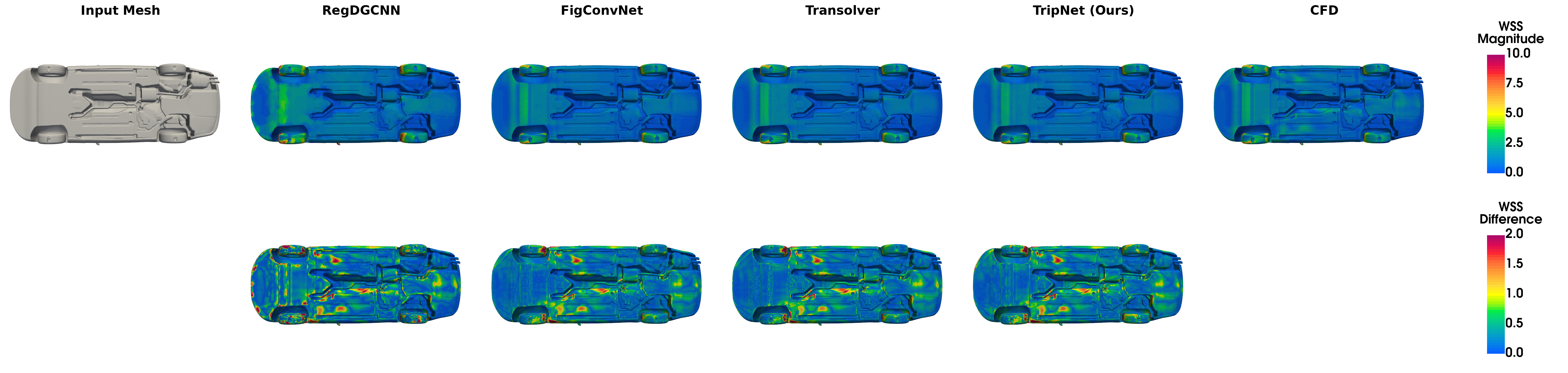}
                \vspace{-4mm}
        \parbox[t]{\textwidth}{\centering (a) Design F\_D\_WM\_WW\_1769 (fastback, detailed underbody, with wheels, and with mirrors).}
        \label{fig:fastback_detailed_underbody_wss}
    \end{subfigure}
    \vspace{3mm}
    \begin{subfigure}
        \centering
        \vspace{0.5mm}
        \includegraphics[width=\textwidth]{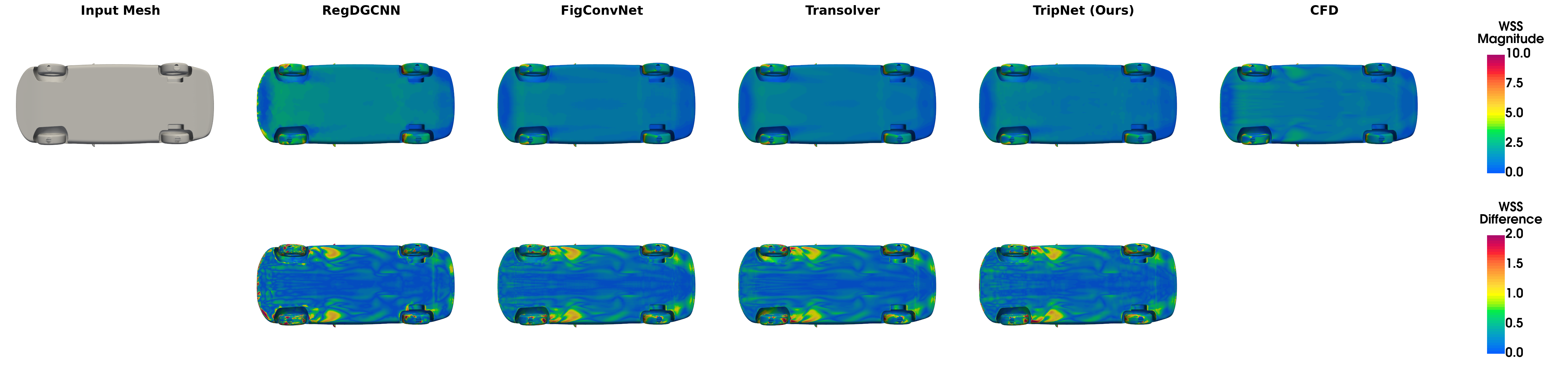}
        \vspace{-4mm}
        \parbox[t]{\textwidth}{\centering (b) Design F\_S\_WWC\_WM\_504 (fastback, smooth underbody, wheels closed, and with mirrors).}
        \label{fig:fastback_smooth_underbody_wss}
    \end{subfigure}
    \vspace{3mm}
    \begin{subfigure}
        \centering
                \vspace{0.5mm}

        \includegraphics[width=\textwidth]{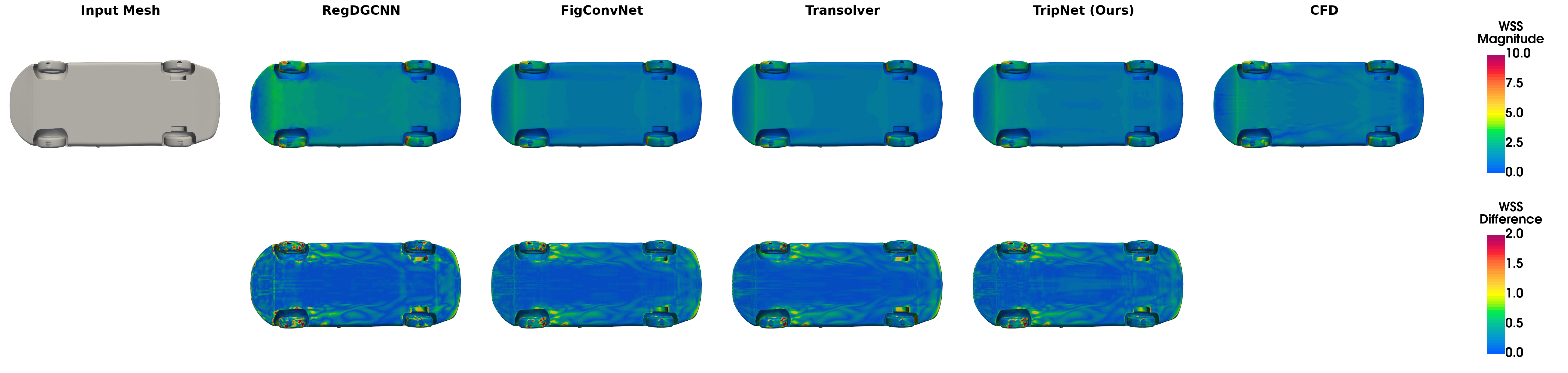}
        \vspace{-4mm}
        \parbox[t]{\textwidth}{\centering (c) Design E\_S\_WWC\_WM\_094 (estateback, smooth underbody, wheels closed, and with mirrors).}
        \label{fig:estateback_underbody_wss}
    \end{subfigure}
    \vspace{2mm}
            \begin{subfigure}
        \centering
                \vspace{0.5mm}

        \includegraphics[width=\textwidth]{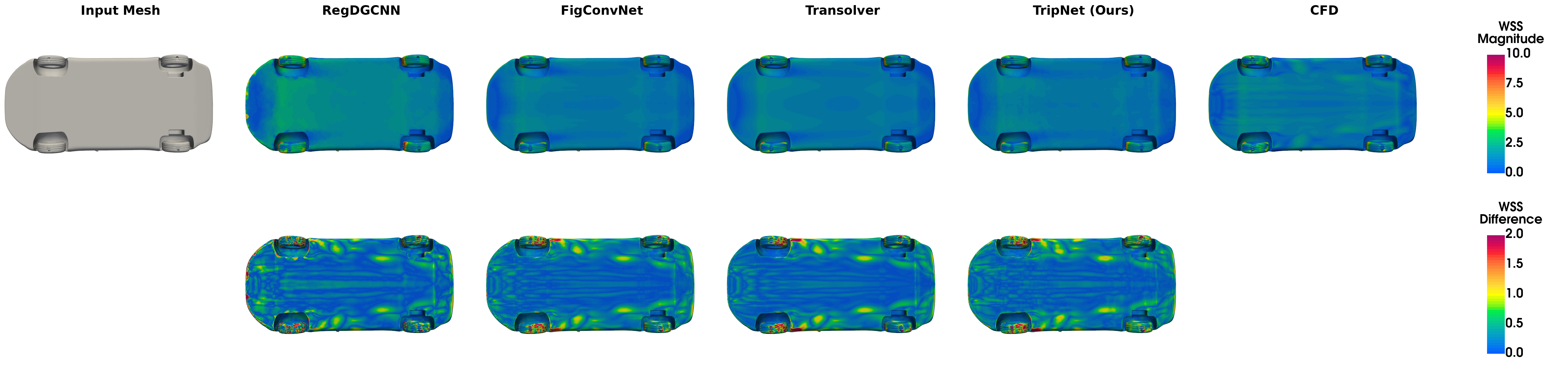}
        \vspace{-4mm}
        \parbox[t]{\textwidth}{\centering (d) Design N\_S\_WW\_WM\_025 (notchback, smooth underbody, with wheels, and with mirrors).}
        \label{fig:notchback_underbody_wss}
    \end{subfigure}
    \vspace{2mm}
    \caption{Comparison of wall shear stress predictions and errors for three car designs from the unseen test set. For each car, the first row shows the input mesh followed by predictions from RegDGCNN, FigConvNet, Transolver, TripNet (ours), and the ground-truth CFD. The second row highlights the absolute difference between the predictions and the ground-truth CFD, visualizing the absolute error distribution for each model.}
    \label{fig:comparison_three_underbody_wss}
\end{figure}

Figures~\ref{fig:comparison_three_isometric_wss} and~\ref{fig:comparison_three_underbody_wss} provide a detailed comparison of wall shear stress predictions and errors across multiple deep learning models for three car designs selected from the unseen test set. These figures highlight differences between model predictions and the ground truth CFD results, visualizing the absolute error distribution for each method. We consider various car categories and configurations, including different underbody and wheel setups, to assess the generalization capabilities of deep learning-based aerodynamic prediction methods across diverse design variations.

\clearpage

\subsection{Volumetric Flow Field}
\label{appendix_3d_flow}

\begin{figure*}[h!]
    \centering
    \includegraphics[width=\linewidth]{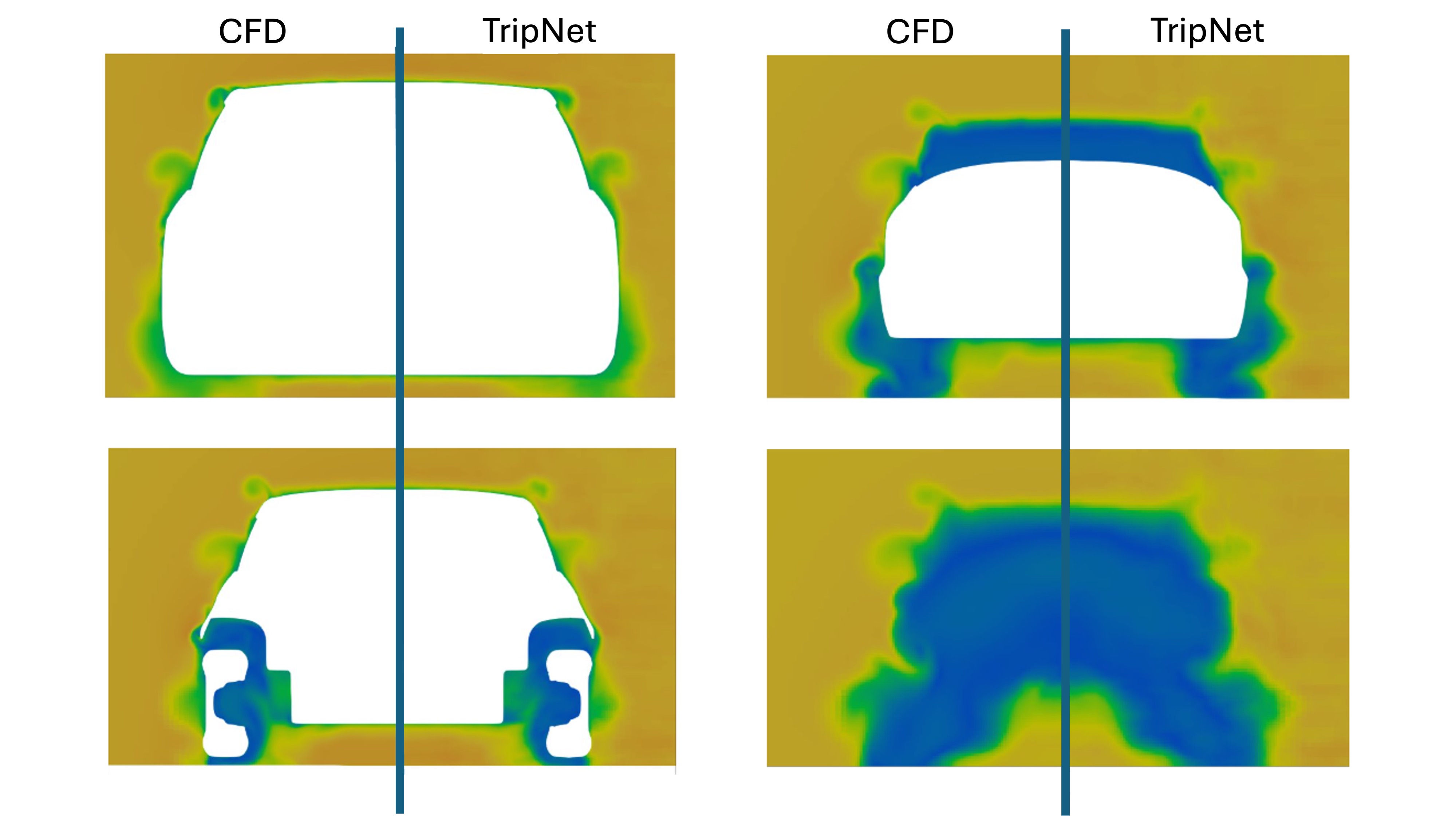}
    \vspace{-5mm}
\caption{Cross-sectional comparison of volumetric field predictions between CFD (left) and TripNet (right) at multiple locations along the car body}
    \label{fig:append_volumetric}
\end{figure*}

As shown in  Figure ~\ref{fig:append_volumetric}, TripNet predictions closely match the CFD ground truth, capturing key flow features near the vehicle surface and in the wake region. The high level of agreement across upstream and downstream sections demonstrates TripNet’s spatial accuracy and generalization capabilities.

\section{Turbulence Modeling}

The derivations and turbulence model discussed in this section are summarized from \cite{wilcox1998turbulence, menter2003ten, menter2021overview, menter1993zonal, spalding1974numerical}, providing a foundation for advanced turbulence modeling approaches capable of addressing the challenges of accurately simulating real-world turbulent flows. 

The Navier-Stokes equations form the cornerstone of fluid dynamics, providing a robust framework for modeling a wide range of fluid phenomena, from simple laminar flows to highly complex turbulent motions encountered in scientific and engineering applications. These equations serve as the mathematical foundation for simulating fluid flow in diverse contexts, including weather systems, ocean currents, and aerodynamic performance of aircraft. Due to their nonlinear nature and the presence of convective acceleration terms, the Navier-Stokes equations are inherently coupled and mathematically complex. Analytical solutions are feasible only for highly simplified cases, making numerical methods and computational fluid dynamics (CFD) indispensable for solving these second-order nonlinear partial differential equations in practical scenarios.

The component-wise representation of the Navier-Stokes equations in Cartesian coordinates is particularly useful in practical applications, as it facilitates detailed modeling of fluid behavior in three-dimensional space. 

\begin{equation}
\frac{\partial u_1}{\partial x_1} + \frac{\partial u_2}{\partial x_2} + \frac{\partial u_3}{\partial x_3} = 0
\end{equation}

\begin{equation}
\begin{aligned}
\left(x_1\right): \quad & \rho\left(\frac{\partial u_1}{\partial t}+u_1 \frac{\partial u_1}{\partial x_1}+u_2 \frac{\partial u_1}{\partial x_2}+u_3 \frac{\partial u_1}{\partial x_3}\right) =-\frac{\partial p}{\partial x_1}+\mu\left[\frac{\partial^2 u_1}{\partial x_1^2}+\frac{\partial^2 u_1}{\partial x_2^2}+\frac{\partial^2 u_1}{\partial x_3^2}\right]+\rho f_1 \\
\left(x_2\right): \quad & \rho\left(\frac{\partial u_2}{\partial t}+u_1 \frac{\partial u_2}{\partial x_1}+u_2 \frac{\partial u_2}{\partial x_2}+u_3 \frac{\partial u_2}{\partial x_3}\right) =-\frac{\partial p}{\partial x_2}+\mu\left[\frac{\partial^2 u_2}{\partial x_1^2}+\frac{\partial^2 u_2}{\partial x_2^2}+\frac{\partial^2 u_2}{\partial x_3^2}\right]+\rho f_2 \\
\left(x_3\right): \quad & \rho\left(\frac{\partial u_3}{\partial t}+u_1 \frac{\partial u_3}{\partial x_1}+u_2 \frac{\partial u_3}{\partial x_2}+u_3 \frac{\partial u_3}{\partial x_3}\right) =-\frac{\partial p}{\partial x_3}+\mu\left[\frac{\partial^2 u_3}{\partial x_1^2}+\frac{\partial^2 u_3}{\partial x_2^2}+\frac{\partial^2 u_3}{\partial x_3^2}\right]+\rho f_3
\end{aligned}
\end{equation}

Here, \( u_1, u_2, u_3 \) represent the velocity components in the \( x_1, x_2, x_3 \) directions, respectively, while \( p \) is the pressure, \( \rho \) is the fluid density, \( \mu \) is the dynamic viscosity, and \( f_1, f_2, f_3 \) are the body force components acting in the respective directions. The terms account for the fluid's inertia, pressure gradient, viscous diffusion, and external forces. For clarity, the
incompressible fluid with constant density and constant viscosity is assumed.

Reynolds averaging is introduced as a key approach to address the complexity of turbulent flows by decomposing the instantaneous flow variables into their mean and fluctuating components. However, this approach leads to the well-known \textit{"closure problem"} due to the additional unknown terms introduced in the Reynolds-averaged equations. To resolve this, turbulence models are required to provide approximations for the unknown terms. These models, employed within the Reynolds-Averaged Navier-Stokes (RANS) framework, enable practical solutions for complex turbulent flows.
\subsection{k-$\omega$ \text{Shear Stress Transport (SST)}}

The $k-\omega$ Shear Stress Transport (SST) model is a widely used two-equation turbulence model that solves for turbulence kinetic energy ($k$) and the specific dissipation rate ($\omega$). The SST model is particularly effective in capturing flow separation, which is critical for accurately predicting aerodynamic and hydrodynamic performance. Both the DrivAerNet~\cite{elrefaie2024drivaernet} and DrivAerNet++~\cite{elrefaie2024drivaernet++} datasets are generated using RANS simulations with the $k-\omega$ SST turbulence model, ensuring high-fidelity aerodynamic predictions.

The combined equations for turbulence kinetic energy (\(k\)) and specific dissipation rate (\(\omega\)) are given by:
\begin{equation}
\frac{D}{D t}
\begin{pmatrix}
\rho k \\
\rho \omega
\end{pmatrix}
=
\nabla \cdot
\begin{pmatrix}
\rho D_k \nabla k \\
\rho D_\omega \nabla \omega
\end{pmatrix}
+
\begin{pmatrix}
\rho G - \frac{2}{3} \rho k (\nabla \cdot \mathbf{u}) - \rho \beta^* \omega k + S_k \\
\frac{\rho \gamma G}{\nu} - \frac{2}{3} \rho \gamma \omega (\nabla \cdot \mathbf{u}) - \rho \beta \omega^2 - \rho \left(F_1 - 1\right) C D_{k\omega} + S_\omega
\end{pmatrix}.
\label{eq:k-omega}
\end{equation}

These equations describe the transport of turbulence kinetic energy (\( k \)) and the specific dissipation rate (\( \omega \)). The terms \( D_k \) and \( D_\omega \) denote the diffusivities of \( k \) and \( \omega \), respectively, while \( G \) represents the generation of turbulence kinetic energy due to mean velocity gradients. The parameters \( \gamma \), \( \beta \), and \( \beta^* \) are model constants governing the dissipation and production of turbulence. The source terms \( S_k \) and \( S_\omega \) account for additional influences on the transport of \( k \) and \( \omega \). These equations are fundamental to the RANS framework, enabling the modeling of turbulence effects in complex fluid flows.

The turbulence viscosity \( \nu_t \) is obtained using the following relation:

\begin{equation}
\nu_t = a_1 \frac{k}{\max \left(a_1 \omega, b_1 F_{23} \mathbf{S} \right)},
\label{eq:turbulence_viscosity}
\end{equation}

where \( a_1 \) and \( b_1 \) are model constants, and \( F_{23} \) is a blending function that ensures a smooth transition between the $k-\omega$ model behavior in the near-wall region and the $k-\varepsilon$ model behavior in the free stream. The term \( \mathbf{S} \) represents the magnitude of the mean rate-of-strain tensor, which plays a crucial role in determining turbulence viscosity \( \nu_t \).

The SST model combines the advantages of the $k-\omega$ and $k-\varepsilon$ turbulence models through a blending function \( F_{23} \), allowing it to perform accurately in both near-wall regions and free-stream conditions. It excels at handling adverse pressure gradients and predicting flow separation, making it highly effective for engineering applications such as aerodynamics and hydrodynamics, where precision in near-wall flow and separation behavior is essential.

\subsection{Total Drag Coefficient}

The total drag coefficient \( C_d \), defined as the sum of the \textbf{pressure drag coefficient} \( c_p \) and the \textbf{shear stress drag coefficient} \( c_f \), can also be expressed in terms of the drag force \( F_d \), dynamic pressure, and reference area as:
\begin{equation}
C_d = c_p + c_f = \frac{F_d}{\frac{1}{2} \rho U^2 A}
\end{equation}
where \( F_d \) is the total drag force, \( \rho \) is the fluid density, \( U \) is the free-stream velocity, and \( A \) is the reference area (typically the frontal projected area).

\paragraph{Pressure Drag Contribution}
The pressure drag coefficient \( c_p \) arises from the pressure distribution on the surface. It can be expressed as:

\begin{equation}
c_p = \text{coeff} \cdot \sum_{i} \left( \mathbf{n}_i \cdot \mathbf{d} \right) \cdot \left( A_i \cdot p_i \right)
\end{equation}
Here, \( \mathbf{n}_i \) is the surface normal vector for the \( i \)-th cell, \( \mathbf{d} \) is the direction of movement (default is \([-1, 0, 0]\)), \( A_i \) is the area of the \( i \)-th surface cell, \( p_i \) is the pressure on the \( i \)-th cell, and \( \text{coeff} = \frac{2}{A \rho U^2} \), where \( \rho \) is the fluid density, \( U \) is the inlet velocity, and \( A \) is the cross-sectional area along the movement direction.


\paragraph{Shear Stress Drag Contribution}
The shear stress drag coefficient \( c_f \) arises from the wall shear stress on the surface. It can also be expressed as:

\begin{equation}
c_f = \text{coeff} \cdot \sum_{i} \left| \mathbf{\tau}_{w,i} \cdot \mathbf{d} \right| \cdot A_i
\end{equation}
Here, \( \mathbf{\tau}_{w,i} \) is the wall shear stress on the \( i \)-th cell, \( \mathbf{d} \) is the direction of movement (default is \([-1, 0, 0]\)), \( A_i \) is the area of the \( i \)-th surface cell, and \( \text{coeff} = \frac{2}{A \rho U^2} \), as described above.


\paragraph{Total Drag Coefficient}
The total drag coefficient \( C_d \) is the sum of the pressure drag and shear stress drag contributions. In both notations:

\begin{equation}
C_d = \text{coeff} \cdot \left[ \sum_{i} \left( \mathbf{n}_i \cdot \mathbf{d} \right) \cdot \left( A_i \cdot p_i \right) + \sum_{i} \left| \mathbf{\tau}_{w,i} \cdot \mathbf{d} \right| \cdot A_i \right]
\end{equation}


Predicting the aerodynamic drag of a car can be approached in two ways: directly estimating the drag coefficient from the 3D geometry or first predicting the surface field quantities, such as pressure and wall shear stress (WSS), and subsequently integrating or summing these fields to compute the drag. The latter approach is particularly valuable as it provides detailed insights into the aerodynamic forces acting on different regions of the car, enabling a more interpretable and physics-informed drag estimation.

\section{Description of the DrivAerNet++ Dataset}
\label{sec:DrivAerNet}
In this study, we utilize the DrivAerNet++ dataset, introduced by \cite{elrefaie2024drivaernet, elrefaie2024drivaernet++}, a comprehensive multimodal dataset specifically developed for aerodynamic car design. It includes over 8,000 unique industry-standard car models, each meticulously modeled using high-fidelity CFD simulations. The dataset features a variety of car body styles, such as fastback, notchback, and estateback, and encompasses various underbody and wheel configurations applicable to both internal combustion and electric cars. 
\begin{table}[h!]
\centering
\normalsize
\caption[Comparison of drag coefficients across different DrivAerNet{\raisebox{0.3ex}{\scriptsize ++}} car configurations.]{Comparison of drag coefficient ($C_d$) ranges and mean values across different car configurations in the DrivAerNet{\raisebox{0.3ex}{\scriptsize ++}} dataset. Each row represents a specific configuration with the corresponding number of samples, $C_d$ range, and mean $C_d$. Each car is normalized by its corresponding projected frontal area. The images illustrate the respective car designs used for the aerodynamic simulations. Two underbody configurations are considered: smooth and detailed, and three wheels/tires configurations: WW (with wheels), WWS (smooth), and WWC (with wheels closed).}
\vspace{2mm}
\resizebox{\textwidth}{!}{%
\begin{tabular}{>{\centering\arraybackslash}p{2cm} @{\,}>{\centering\arraybackslash}p{2cm} @{\,}>{\centering\arraybackslash}p{1.5cm} @{\,}>{\centering\arraybackslash}p{7cm} @{\,}>{\centering\arraybackslash}p{1.5cm} @{\,}>{\centering\arraybackslash}p{3cm} @{\,}>{\centering\arraybackslash}p{2cm}}
\hline
\raisebox{-1.5ex}{Configuration} & \raisebox{-1.5ex}{Underbody} & \raisebox{-1.5ex}{Wheels} & \raisebox{-1.5ex}{Car Design} & Number of Samples & \raisebox{-1.5ex}{[min $C_d$, max $C_d$]} & \raisebox{-1.5ex}{Mean $C_d$}  \\
\hline
\vspace{0.1cm} \\ 
Estateback & Smooth & WW & \adjustbox{valign=m}{\includegraphics[width=0.35\textwidth]{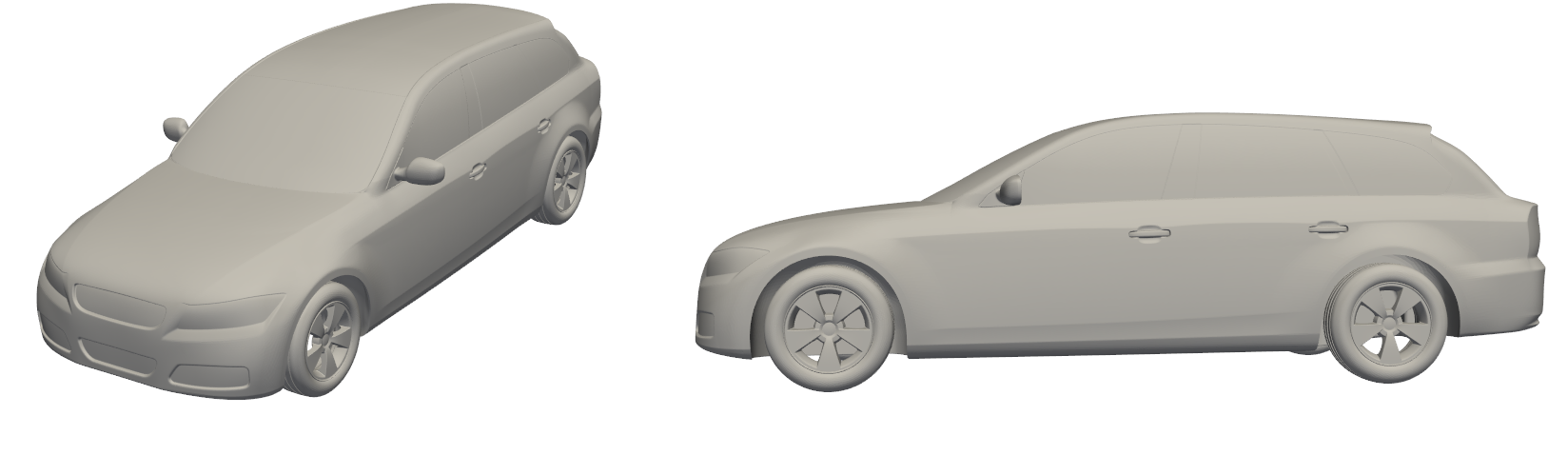}} & 698 & [0.2223, 0.3198] & 0.2725  \\ 
Estateback & Smooth & WWC & \adjustbox{valign=m}{\includegraphics[width=0.35\textwidth]{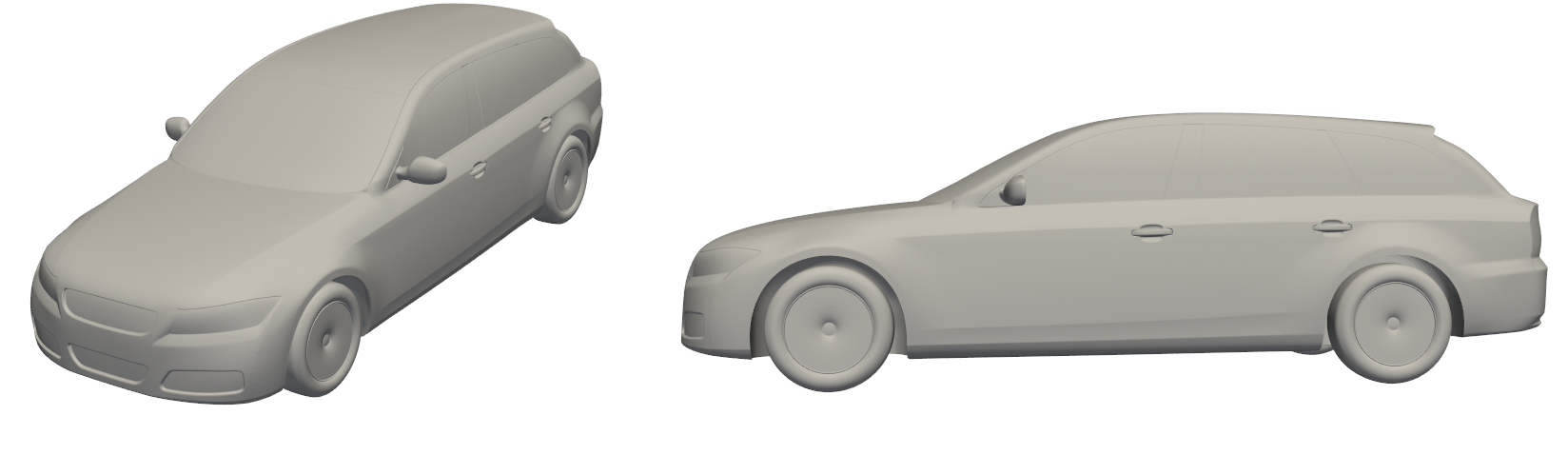}} & 688 & [0.2192, 0.3164] & 0.2706  \\ 
Notchback & Smooth & WW & \adjustbox{valign=m}{\includegraphics[width=0.35\textwidth]{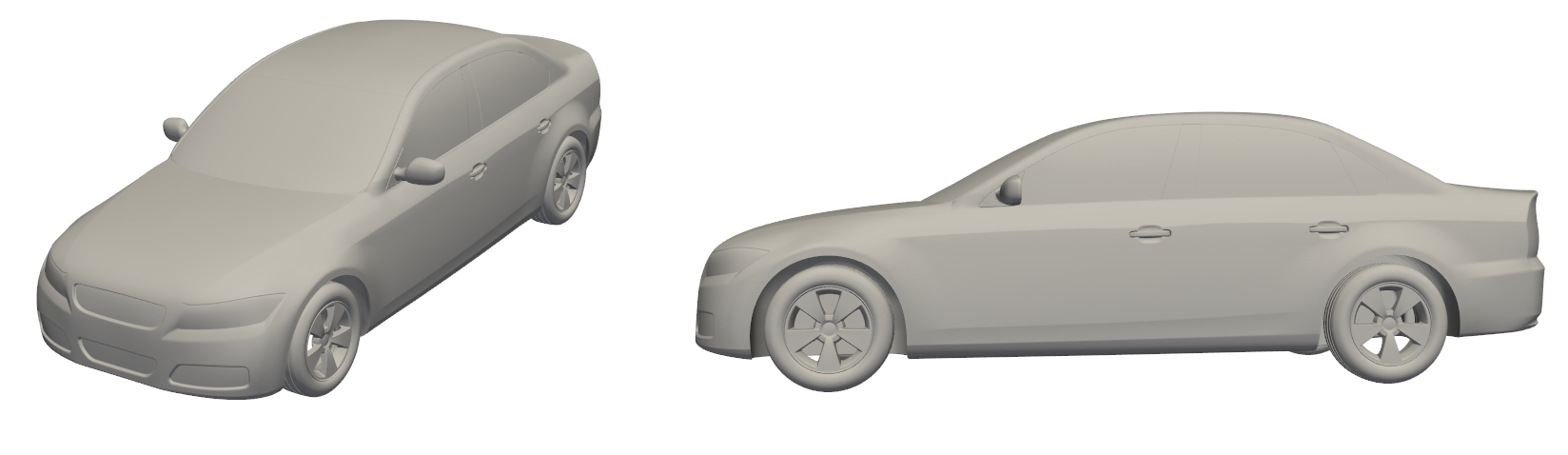}} & 676 & [0.2012, 0.2978] & 0.2457 \\ 
Notchback & Smooth & WWS & \adjustbox{valign=m}{\includegraphics[width=0.35\textwidth]{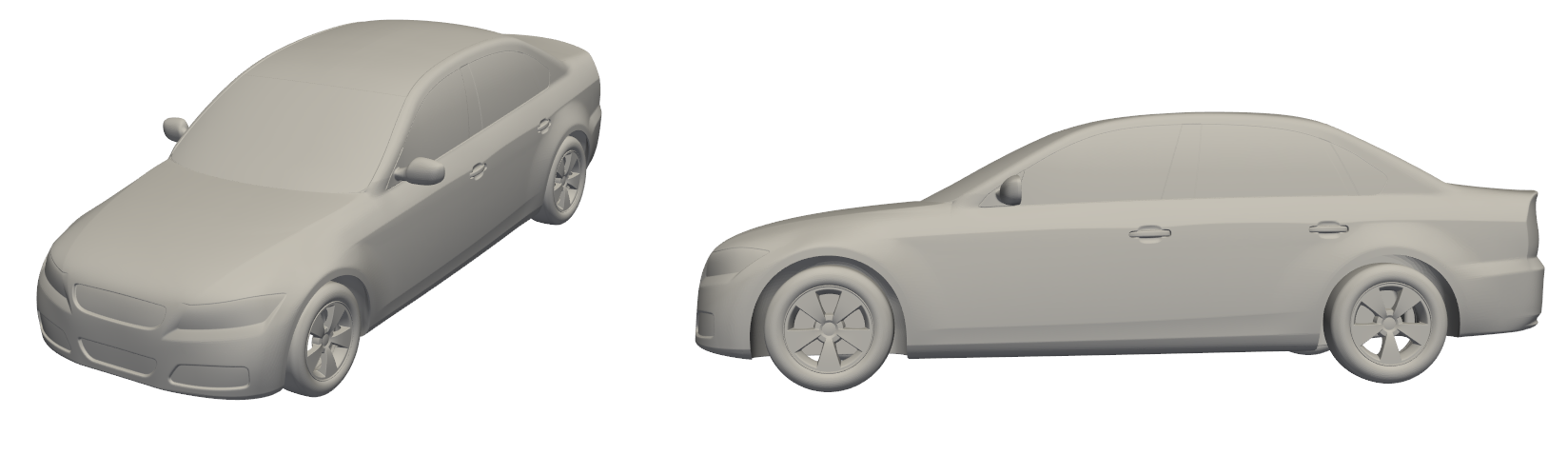}} & 341 & [0.2049, 0.2985] & 0.2443 \\ 
Notchback & Smooth & WWC & \adjustbox{valign=m}{\includegraphics[width=0.35\textwidth]{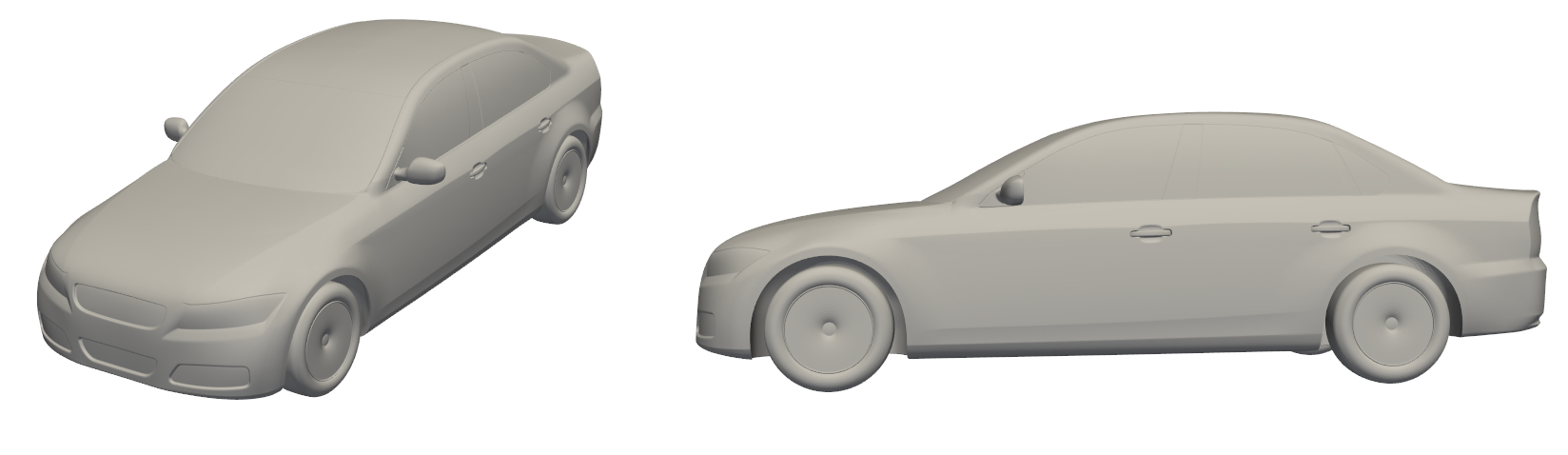}} & 386 & [0.1998, 0.2969] & 0.2473  \\ 
Fastback & Smooth & WWS & \adjustbox{valign=m}{\includegraphics[width=0.35\textwidth]{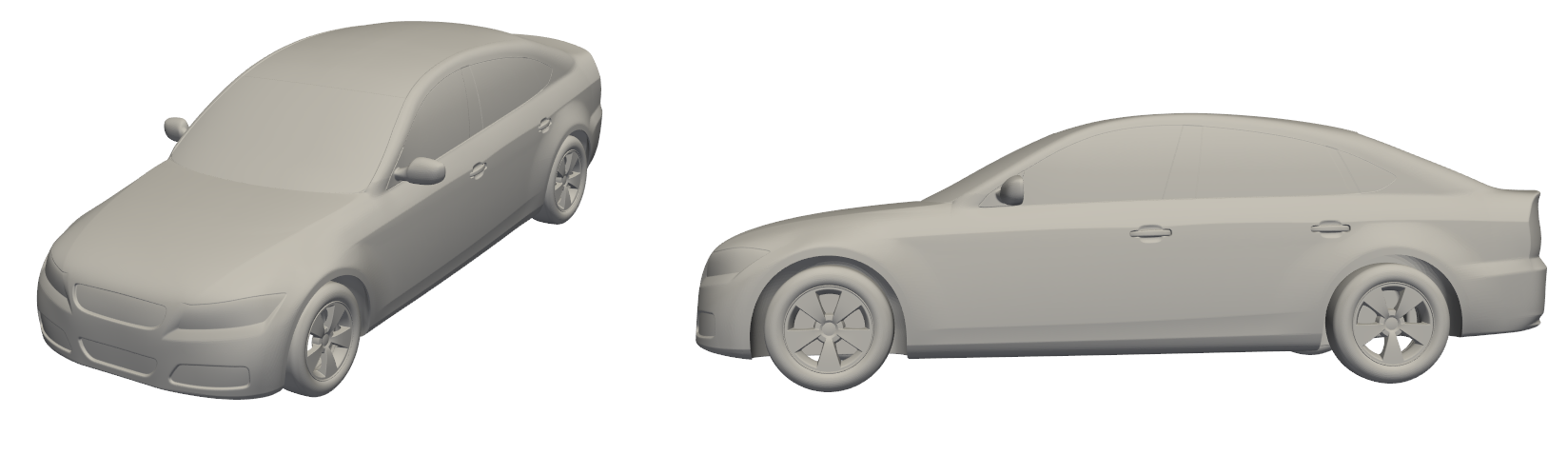}} & 684 & [0.2037, 0.2901] & 0.2417 \\ 
Fastback & Smooth & WWC & \adjustbox{valign=m}{\includegraphics[width=0.35\textwidth]{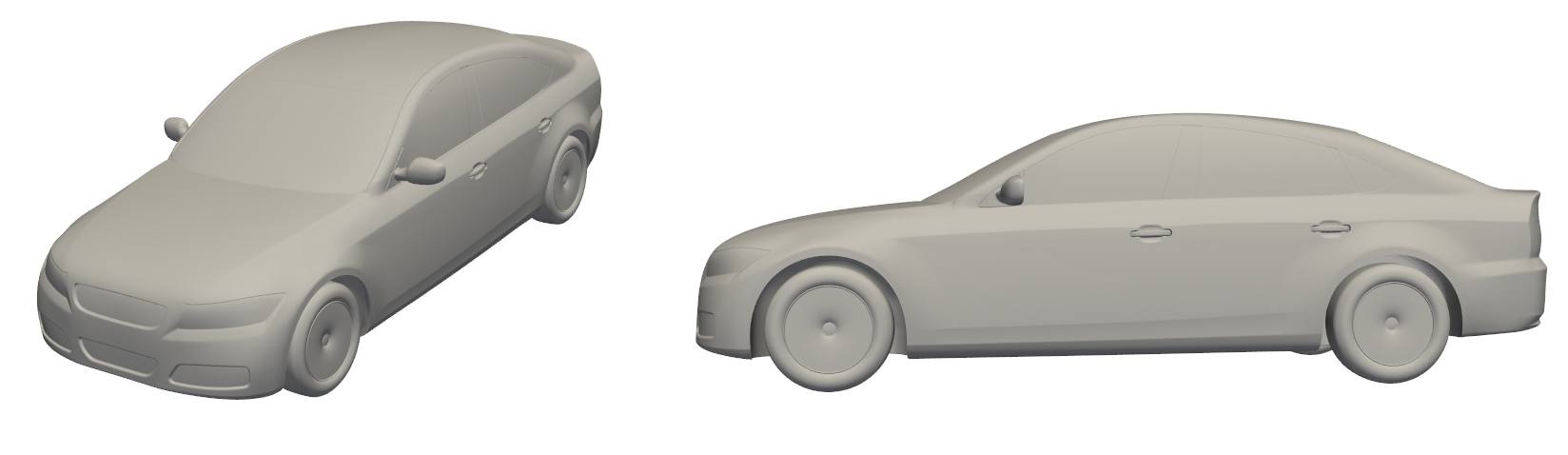}} & 692 & [0.2009, 0.2971] & 0.2463 \\ 
\hline
\vspace{0.1cm} \\
Fastback & Detailed & WW & \adjustbox{valign=m}{\includegraphics[width=0.35\textwidth]{Imgs/F_S_WWS_WM_7_F_S_WWS_WM_7_002_combined.png}} & 3956 & [0.2372, 0.3194] & 0.2732  \\ \hline
\end{tabular}%
}
\label{tab:car_design}
\end{table}

Accompanying each model are detailed 3D meshes, parametric models, aerodynamic coefficients, as well as extensive flow and surface field data. DrivAerNet++ also provides segmented parts for car classification and point cloud data, enabling a broad spectrum of machine learning applications including design optimization, generative modeling, surrogate model training, CFD simulation acceleration, and geometric classification. This dataset represents a significant advancement over existing datasets due to its size, realistic industry-standard car shapes, shape diversity, and higher fidelity of its CFD simulations, making it an invaluable resource for training deep learning models. Table~\ref{tab:car_design} provides a detailed comparison of drag coefficient ($C_d$) ranges, mean values, and sample sizes across various DrivAerNet++ car configurations.





\section{Evaluation Metrics}

The \textbf{Mean Squared Error (MSE)} measures the average squared difference between predicted and true values, penalizing larger errors more heavily, making it sensitive to outliers:

\begin{equation}
\text{MSE} = \frac{1}{N} \sum_{i=1}^{N} (y_i - \hat{y}_i)^2
\end{equation}

where \( y_i \) and \( \hat{y}_i \) represent the ground truth and predicted values, respectively, and \( N \) is the total number of samples.

The \textbf{Mean Absolute Error (MAE)} calculates the average absolute difference between predictions and true values, providing a more robust measure of error that is less sensitive to outliers:

\begin{equation}
\text{MAE} = \frac{1}{N} \sum_{i=1}^{N} |y_i - \hat{y}_i|
\end{equation}

The \textbf{Maximum Absolute Error (Max AE)} quantifies the worst-case prediction error by identifying the largest absolute difference between predicted and true values:

\begin{equation}
\text{Max AE} = \max \big(|y_i - \hat{y}_i|\big)
\end{equation}

The \textbf{Coefficient of Determination (\( R^2 \))} measures how well the predicted values explain the variance of the ground truth values. It is defined as:

\begin{equation}
R^2 = 1 - \frac{\sum_{i=1}^{N} (y_i - \hat{y}_i)^2}{\sum_{i=1}^{N} (y_i - \bar{y})^2}
\end{equation}

where \( \bar{y} \) is the mean of the ground truth values. 

The \textbf{Relative \( L_2 \) Error} evaluates the normalized Euclidean difference between predictions and true values, measuring model performance relative to the magnitude of the true values:

\begin{equation}
\text{Relative } L_2 \text{ Error} = \frac{\| \hat{\mathbf{y}} - \mathbf{y} \|_2}{\| \mathbf{y} \|_2}
\end{equation}

where \( \|\cdot\|_2 \) represents the Euclidean norm.

Similarly, the \textbf{Relative \( L_1 \) Error} computes the normalized sum of absolute differences between predictions and true values, relative to the sum of absolute true values:

\begin{equation}
\text{Relative } L_1 \text{ Error}= \frac{\| \hat{\mathbf{y}} - \mathbf{y} \|_1}{\| \mathbf{y} \|_1}
\end{equation}

where \( \|\cdot\|_1 \) denotes the L1 norm, representing the sum of absolute values.

\paragraph{Evaluation Strategy} 

For surface field predictions (pressure and wall shear stress), we prioritize evaluating models based on the relative \( L_1 \) and \( L_2 \) errors. For drag coefficient estimation (a scalar prediction task), models are ranked according to their \( R^2 \) performance. Meanwhile, volumetric flow field prediction performance is assessed using the mean squared error (MSE). These metrics collectively provide a comprehensive evaluation of accuracy, robustness, and worst-case performance across different aerodynamic prediction tasks.

\end{document}